\documentclass[a4paper,prd,aps,preprint,tightenlines,superscriptaddress,nofootinbib,showpacs,showkeys,floatfix]{revtex4}
\usepackage{mathrsfs}
\usepackage{amsfonts}
\usepackage{array}
\usepackage{epsfig}
\usepackage{rotating}
\usepackage{multirow}

\usepackage{amsmath}    
\usepackage{verbatim}   
\usepackage{color}      
\usepackage{colordvi}
\usepackage{bm}

\usepackage{subfigure}  
\usepackage{url}
\usepackage[colorlinks = true,
            linkcolor = blue,
            urlcolor  = blue,
            citecolor = blue,
            anchorcolor = blue]{hyperref}
\usepackage{doi}

\usepackage{pstricks,rotating, graphicx}
\usepackage{cancel}

\usepackage[normalem]{ulem}

\usepackage{tablefootnote}  

\newcommand{\GeV}{\ensuremath{\,\mathrm{GeV}}}

\newcommand{\msbar}{$\overline{\mathrm{MS}}\, $}

\graphicspath{{./figs/}}

\usepackage{pslatex}
\usepackage{txfonts}
\usepackage[latin1]{inputenc}
\usepackage[T1]{fontenc}

\begin{document}
\begin{titlepage}
\thispagestyle{empty}
\noindent
DESY 24-092
\hfill
December 2024 \\
\vspace{1.0cm}

\begin{center}
  {\bf \Large
    NNLO PDFs driven by
    top-quark data 
  }
  \vspace{1.25cm}

 {\large
   S.~Alekhin$^{\, a}$,
   M. V. Garzelli$^{\, a}$,
   S.-O.~Moch$^{\, a}$
   and
   O. Zenaiev$^{\, a}$
 }
 \\
 \vspace{1.25cm}
 {\it
   $^a$ 
II. Institut f\"ur Theoretische Physik, Universit\"at Hamburg \\
   Luruper Chaussee 149, D--22761 Hamburg, Germany \\
 }

\vspace{1.4cm}
\large {\bf Abstract}
\vspace{-0.2cm}
\end{center}
We study the impact of state-of-the-art top-quark data collected at the Large Hadron Collider on parton distribution functions (PDFs). 
Following the ABMP methodology, the fit extracts simultaneously proton PDFs, 
the strong coupling $\alpha_s(M_Z)$ and heavy-quark masses at next-to-next-to-leading order (NNLO) accuracy in QCD.
It includes recent high-statistics data on absolute total inclusive cross sections for $t\bar{t}+X$, the sum of $(t + X)$ and  $(\bar{t} + X)$  hadroproduction, and normalized inclusive data double-differential in the invariant mass and rapidity of the $t\bar{t}$ pair at $\sqrt{S}=13$~TeV. 
The gluon PDF at large $x$ and the top-quark mass value derived from these data are well compatible with the previous ABMP16 results, but with significantly smaller uncertainties, reduced by up to a factor of two.
At NNLO in QCD we obtain for the strong coupling the value 
$\alpha_s^{(n_f=5)}(M_Z)= 0.1150 \pm 0.0009$
and for the top-quark mass in the ${\overline{\mbox{MS}}}$-scheme
$m_t(m_t) = 160.6 \pm 0.6$~GeV, 
corresponding to $m_t^{\rm pole} = 170.2 \pm 0.7$~GeV 
in the on-shell scheme.
The new fit, dubbed ABMPtt, is publicly released in grids in LHAPDF format.

\end{titlepage}

\newpage
\setcounter{footnote}{0}
\setcounter{page}{1}

\section{Introduction}
\label{sec:intro}

The experimental collaborations at the Large Hadron Collider (LHC) have done a tremendous work to produce analyses of top-quark events
with unprecedented precision and accuracy at increasing integrated luminosity. In particular, in case of $t\bar{t} +X$ hadroproduction - the process, among those involving top quarks, with the highest cross section - 
first results on total cross sections at a center-of-mass energy $\sqrt{S}=$~13.6~TeV
have emerged from Run~3 data~\cite{CMS:2023qyl,ATLAS:2023slx}, whereas Run~2, besides total cross sections~\cite{ATLAS:2019hau, Khachatryan:2016kzg, Sirunyan:2018goh, CMS:2016rtp, ta13lj,ATLAS:2023gsl} and single-differential ones~\cite{ATLAS:2019hxz,a200609274,top20001,ATLAS:2023gsl}, 
has provided high-quality cross-section data double differential~\cite{top20001, a190807305, a200609274} and triple differential~\cite{top18004, CMS:2024ybg} in the invariant mass $M(t\bar{t})$ and the ra\-pi\-di\-ty $y(t\bar{t})$ of the $t\bar{t}$ pair, as well as in the number of additional jets accompanying the pair, at $\sqrt{S}=13$~TeV. This has followed the first double-differential data produced during Run~1 at $\sqrt{S}=8$~TeV~\cite{top14013}, characterized indeed by a lower accuracy, and further analyses during Run 1 producing both
total cross sections at
$\sqrt{S}=7$~TeV~\cite{Aad:2014kva,     Aad:2014jra,            Khachatryan:2016mqs, Khachatryan:2016yzq, ATLAS:2012gpa, Aad:2015dya, Chatrchyan:2012vs,  Chatrchyan:2013kff, Chatrchyan:2013ual,   Aad:2012vip} and 8~TeV~\cite{Aad:2014kva, Khachatryan:2016mqs, Khachatryan:2016yzq,  Aad:2015pga,   Khachatryan:2014loa, Aaboud:2017rgh,   Khachatryan:2015fwh} and single-differential distributions at $\sqrt{S}=7$~TeV~\cite{a160707281, a14070371} and 8~TeV~\cite{CMS:2015rld,a151104716, a160707281}.
Additionally, special low intensity runs conducted in 2015 and 2017 led to total cross section measurements at $\sqrt{S}=5.02$~TeV~\cite{CMS:2017zpm, tc5, ta5, CMS:2024zbz}. Further information and 
references, especially as for CMS analyses, can be found in the recent CMS review on top-quark mass measurements~\cite{CMS:2024irj}.
  
The existence of the aforementioned multi-differential
data is indeed crucial for the simultaneous fit of parton distribution functions (PDFs), the top-quark mass $m_t$ and the strong coupling $\alpha_s$. In particular, the shape of the $y(t\bar{t})$ distributions is especially sensitive to the gluon PDF, the shape of the $M(t\bar{t})$ distributions is particularly
sensitive to the top-quark mass~\footnote{At leading order, the longitudinal momentum fractions $x_1$ and $x_2$ of the 
partons undergoing an hard scattering from the colliding protons 
are related to $y(t\bar{t})$ and $M(t\bar{t})$ by $x_{1,2}=\frac{M(t\bar{t})}{\sqrt{S}}\mathrm{exp}(\pm y(t\bar{t}))$. 
Ref.~\cite{Czakon:2016vfr}, considering NNLO calculations, has shown that the shape of the $M({t\bar{t}})$ distribution is very sensitive to changes in the $m_t$ value, whereas $y_t$ and $y({t\bar{t}})$ are less sensitive to it.}, 
whereas the shape of the distribution of the number of additional jets is directly sensitive to the $\alpha_s(M_Z)$ value.

In this work, we provide the first simultaneous fit of PDFs, $m_t(m_t)$ and $\alpha_s(M_Z)$ at next-to-next-to-leading order (NNLO) in the \msbar scheme using double-differential data, following the next-to-leading (NLO) fits already published by the CMS experimental collaboration in Ref.~\cite{top18004} and by us
in our previous work~\cite{Garzelli:2020fmd} and the approximate NNLO fit by Ref.~\cite{Guzzi:2014wia}. 
As a basis we rely on the ABMP16 fit methodology, extensively described in Ref.~\cite{Alekhin:2017kpj}. Even 
the latter fit extracted simultaneously the
PDFs, $m_t(m_t)$ and $\alpha_s(M_Z)$, however without using any differential $t\bar{t} + X$ cross-section data. We do not use any NNLO/NLO $K$-factor approximation, but compute the predictions for our fit at NNLO including QCD radiative corrections,
as detailed in the following section. 

Other PDF collaborations have also incorporated part of the existing double-differential data into their PDF fits: besides the already mentioned Ref.~\cite{Guzzi:2014wia}, Ref.~\cite{Czakon:2019yrx}  examined the impact of double-differential $t\bar{t}+X$ data at $\sqrt{S}=8$~TeV on the gluon distribution, using as a basis the CT14HERA2 global PDF fit, 
and the fits by CT18~\cite{Hou:2019efy}, MSHT20~\cite{Bailey:2020ooq} and
NNPDF4.0~\cite{NNPDF:2021njg} also made use of the same dataset.
We emphasize however that, according to our knowledge, so far these fits are limited to the use of double-differential Run~1 data. 
Additionally, the NNPDF4.0 collaboration has included single-differential data from Run~2 in their fits following Ref.~\cite{Czakon:2016olj}, where an extended NNPDF3.0 PDF fit, incorporating single-differential $t\bar{t}+X$ data at $\sqrt{S}=$8~TeV, was performed~\footnote{The NNPDF3.1 PDF fit~\cite{NNPDF:2017mvq} has
also incorporated   $t\bar{t}+X$ datasets  from Run~1, 
single differential in $y({t\bar{t}})$~\cite{ CMS:2015rld,   a151104716}.}.
More recently, they have evaluated the impact of a more updated and general set of top-quark related data both on SM- and SMEFT-PDF fits~\cite{Kassabov:2023hbm}, with special emphasis on full-luminosity Run II datasets available at the time of their work, including, among others, some double-differential datasets.  
The CT18, MMHT and MSHT20 collaborations have also made use of single-differential data at $\sqrt{S} = 8$~TeV from Run~1. A specific study using single-differential distributions from ATLAS and CMS in the MMHT framework has been conducted in Ref.~\cite{Bailey:2019yze}, showing some tension.  
An extended CT18 fit, including single-differential $t\bar{t}+X$ data at $\sqrt{S}=$~13~TeV, has been recently presented in Ref.~\cite{Ablat:2023tiy}.
On the other hand, a fit of top-quark mass and $\alpha_s(M_Z)$, using various single-differential distributions at $\sqrt{S}=$8 TeV, for fixed PDFs, have also been performed~\cite{Cooper-Sarkar:2020twv}, followed by a recent work in the framework of the MSHT20 PDF fit~\cite{Cridge:2023ztj}, that, using a subset of the top-quark datasets in MSHT20, has found a best-fit $m_t^{\rm pole}$ value compatible with the PDG~\cite{ParticleDataGroup:2024cfk}. This work has additionally analyzed the impact of different top-quark mass values on gluon PDFs, for fixed $\alpha_s(M_Z)=0.118$, finding that increasing values of $m_t^{\rm pole}$ 
  lead to an upwards shift of the central gluon at large $x$, 
thus accounting for relevant correlations.
   None of these studies so far 
has published
a simultaneous fit of PDFs, $\alpha_s(M_Z)$ and $m_t(m_t)$.

In the original ABMP16 fit~\cite{Alekhin:2017kpj} only total cross-section data for $t\bar{t} + X$ and single-top production available at the time of the fit (2016) were considered.
In this work, we update also this part of the fit, by considering recent $t\bar{t} + X$ total cross-section data, with decreased uncertainties with respect to the past.  Recent total
single-top cross-section data at the LHC~\cite{ATLAS:2019hhu, CMS:2018lgn,  ATLAS:2023hul}, summing up $t$ and $\bar{t}$ events, are also included, together with the Tevatron combinations published in Ref.~\cite{CDF:2015gsg}. 
For completeness, we mention that single-top production data have also been incorporated for the first time in the NNPDF4.0 fit. 
They have analyzed $t$-channel production at the LHC, considering both total cross sections and $y({t, \bar{t}})$ distributions, and included NNLO corrections to their calculations by using NNLO/NLO $K$-factors.
On the other hand, the MSHT20 collaboration has not included single-top production data, due to low statistics and moderate impact in their fit, but has checked that the datasets available at the time of their fit were consistent with their predictions.

This manuscript is organized as follows. 
In Section~\ref{sec:input} we discuss the input for the ABMPtt fit, reviewing 
the theory predictions and the experimental data used and providing some details on the fit methodology and the statistical analysis.
Section~\ref{sec:res} contains the results of different variants of the fit, comparing them 
to the original ABMP16 PDFs as well as to those of other groups. 
We present benchmarks with respect to the measured $t\bar{t} +X$ double-differential cross sections and discuss the value of
$m_t(m_t)$ extracted from the fit.
In Section~\ref{sec:conclu} we draw our conclusions. 
Some details on the ABMPtt PDF eigenvectors are reported in the Appendix~\ref{appA:table} 
and further details on various versions of the ABMP PDFs are illustrated in Appendix~\ref{appB:allpdfs}.

\section{Input for the $\textrm{ABMPtt}$ fit}
\label{sec:input}

In this section we describe the input used for the ABMPtt fit of PDFs, $\alpha_s(M_Z)$ and $m_t(m_t)$ in terms of theory predictions, experimental data and methodology.
We focus only on top-quark related processes, keeping the input for the other processes considered in the fit 
(world data on deep-inelastic-scattering (DIS), Drell-Yan (DY) production at colliders and in fixed-target experiments, etc.), 
both in terms of theory predictions and experimental data, identical to the one used in the ABMP16 case of Ref.~\cite{Alekhin:2017kpj}.
Although updated data exist for some of these processes, e.g. DY production, we choose  to stick to the set of data already used in the past to illustrate exclusively the differences, which come from extending the fit to the analysis of top-quark data. In the most recent version of the ABMP16 fits without and with top-quark data, dubbed respectively ABMP16new and ABMPtt, we make however a more consistent use of the DY data with respect to the published version of ABMP16, by applying a more refined fit methodology, as better explained in subsection~\ref{sec:metho}.

\subsection{Theory predictions for top-quark related processes}
\label{sec:theo}
Theory predictions for $t\bar{t} + X$ hadroproduction at NNLO are available since long 
time
in case of total cross sections, whereas have become available more recently in case of differential distributions, using different techniques for the subtraction of infrared divergences in the combination of double-virtual, double-real and real-virtual contributions.
In particular, we have made use of the code {\texttt{HATHOR}}~\cite{Aliev:2010zk} and {\texttt{Fasttop}}~\cite{Moch:unpublished} 
for computing the total cross sections, as was done already in the ABMP16 fit.
We have cross-checked the results with those of the \texttt{MATRIX} code~\cite{Grazzini:2017mhc}, published after the ABMP16 fit, finding agreement. 
In the case of single-top production, for which we consider only total cross sections, we make use of an extended version~\cite{Kant:2014oha} of the code \texttt{HATHOR} for computing our predictions. 
The NNLO QCD corrections to single-top production~\cite{Brucherseifer:2014ama, Berger:2016oht, Campbell:2020fhf} have been treated as in the ABMP16 fit, see Ref.~\cite{Alekhin:2017kpj} for a discussion~\footnote{Single-top hadroproduction has not been implemented in~\texttt{MATRIX} yet.}.

For the double-differential  $t\bar{t} + X$ distributions, on the other hand, we used a customized version of {\texttt{MATRIX}} interfaced to {\texttt{PineAPPL}}~\cite{Carrazza:2020gss}. 
More details on this version and on the interface can be found in our previous work~\cite{Garzelli:2023rvx}. Here we limit ourselves to mention that we cross-checked the results of {\texttt{MATRIX}}, based on $q_T$-subtraction for the treatment of IR divergences~\cite{Catani:2007vq, Bonciani:2015sha, Catani:2019iny, Catani:2019hip}, with those published in Ref.~\cite{Czakon:2016dgf} and obtainable from the {\texttt{HighTEA}} platform~\cite{Czakon:2023hls}, based on the {\texttt{STRIPPER}} approach~\cite{Czakon:2014oma, Czakon:2011ve, Czakon:2010td} and the same numerical two-loop amplitudes~\cite{Czakon:2008zk, Czakon:2013goa}, finding agreement within 1\%. 
We also observe that the bin-by-bin uncertainties associated with the latter predictions are quoted to be 1\%~\cite{Czakon:2016dgf, Catani:2019hip}. 
We recall also that $q_T$-subtraction, differently from {\texttt{STRIPPER}}, is a 
non-local approach and that approximate cross sections are computed in practice 
introducing a cut-off $r_0 = q_{T,min}/M({t\bar{t}})$ on the dimensional quantity
$r = q_T/M({t\bar{t}})$, using a finite small value of $r_0$, which acts effectively as a slicing parameter.
The ``true'' cross sections correspond to the limit $r_{0} \rightarrow 0$. 
In the version of the {\texttt{MATRIX}} code that we used, developed from the code used for producing the predictions in
   Ref.~\cite{Catani:2019hip} and specifically tailored to $t\bar{t}+X$
production, 
this limit is automatized and performed in practice only in the case of the total cross sections, after computing them for a number of $r_0$ values. We have checked that, varying $r_0$ within a suitable range of values, saturation for $r_0 \rightarrow 0$ is reached even in the case of differential distributions. The theory predictions for double-differential distributions used in the following were all obtained from runs with $r_0 = 0.0015$, after having checked that they agree bin-by-bin
with those for $r_0 = 0.0005$ within a few per mille.  
Therefore, while performing the fit in this work, we have neglected the theory uncertainty related to the bin-by-bin extrapolation to $r_0 \rightarrow 0$. 

The central renormalization and factorization scales are fixed to $\mu_R=\mu_F=H_T/4$ in the case of double-differential distributions, with $H_T$ given by the sum of the transverse masses of the top and antitop quarks $\sqrt{p_{T,t}^2+m_t^2} + \sqrt{p_{T,\bar{t}}^2+m_t^2}$, whereas $\mu_R=\mu_F=m_t$ for total cross sections. 
Uncertainties related to scale variations are not accounted for in this analysis.
The inclusion of scale variation uncertainty in PDF fits is presently a topic of intense debate (see e.g. recent works by the MSHT and NNPDF collaborations~\cite{Harland-Lang:2018bxd, McGowan:2022nag,   NNPDF:2024dpb}), whereas our study on the effect of scale uncertainties in NNLO fits of top-quark mass at fixed PDFs has shown that the latter are quite moderate when including double-differential distributions in the fits, e.g. amounting to $\sim 0.2$~GeV for the extracted value of $m_t^{\rm pole}$~\cite{Garzelli:2023rvx}.

The tables of double-differential $t\bar{t}+X$ cross sections  were generated by {\texttt{MATRIX+PineAPPL}}
using as input the top-quark mass $m_t^{\rm pole}$ renormalized in the on-shell scheme, as customary. 
The top-quark mass was then converted to the \msbar scheme, using conversion formulas at three-loops, and $m_t(m_t)$ was used during our fit procedure. 
More precisely, one could have even computed cross sections directly using as input $m_t(m_t)$, as done for instance in Refs.~\cite{Dowling:2013baa,Catani:2020tko}. The latter procedure, however, is much more involved and, considering the need to repeat computations for many different top-quark mass values, would have required a prohibitive CPU time. The dif\-fe\-ren\-ces between cross sections using our approximation and the latter more rigorous methodology are of higher order.

\subsection{Experimental data for $t\bar{t}+X$ and single-top hadroproduction}
\label{sec:exp}

First of all, the ABMPtt fit includes updated results for the absolute total
cross sections for $t\bar{t}+X$ hadroproduction at the Tevatron and the LHC.
At this aim, among the many datasets a\-vai\-la\-ble,
we choose to use the 
datasets and/or combinations also selected  by the LHC Top Working Group as of fall 2023~\footnote{An updated selection has recently  been made available by the LHC Top Working Group (spring 2024). However, the modifications have in practice 
only marginal consequences  for the work reported in this manuscript.}. 
These data span a wide interval of $\sqrt{S}$ values,
ranging from 5.02~TeV to 13.6~TeV. 
The used datasets~\cite{CDF:2013hmv,tc5,ta5,     ATLAS:2022aof,     Sirunyan:2018goh,     top20001,  ta13lj,  ATLAS:2023gsl,     CMS:2023qyl,ATLAS:2023slx} 
are listed in the website of the LHC Top Working Group~\cite{lhctopwg}.

\begin{figure}[h!]
  \includegraphics[width=0.62\textwidth]{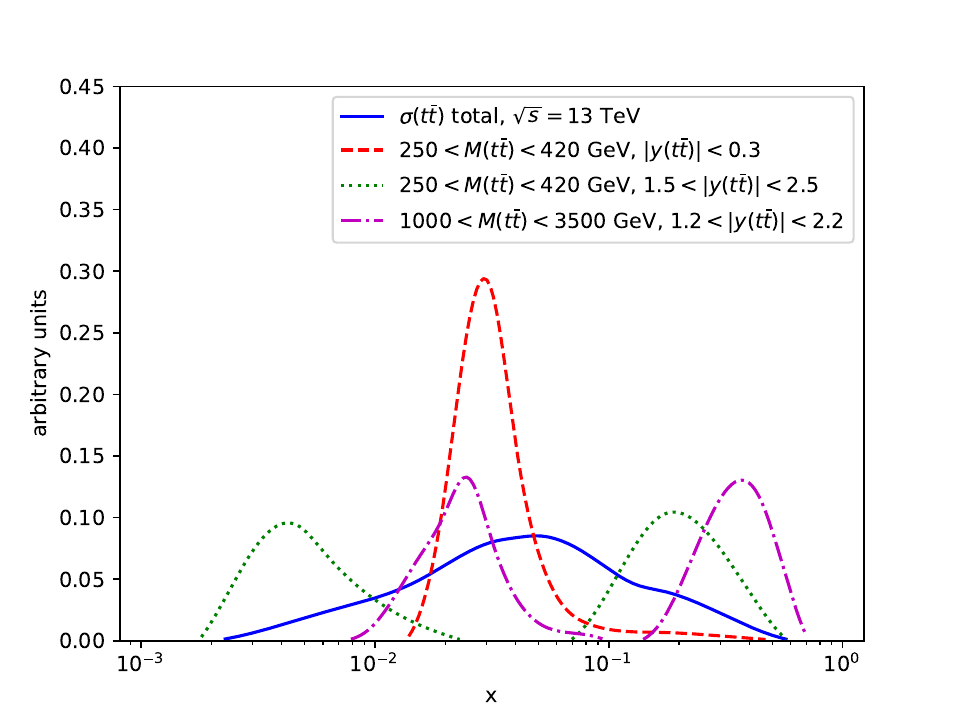}
  \caption{\label{fig:xparton}
    NNLO predictions for $t\bar{t}+X$ production at $\sqrt{S}=13$~TeV as a function of $x$, for different kinematic ranges of $M(t\bar{t})$ and $y(t\bar{t})$, corresponding to the double-differential analysis of Ref.~\cite{top20001}.
    For illustration purposes, all predictions are normalized to unity.}
\end{figure}

Second, the fit includes data on single-top hadroproduction in the $s$- and $t$-channel from the Tevatron~\cite{CDF:2015gsg} and in the $t$-channel from the LHC. 
In particular, we consider the recent combinations of the ATLAS and CMS experiments for $\sqrt{S}$~=~7 and 8 TeV~\cite{ATLAS:2019hhu}.
  Additionally, we consider the separate data points by ATLAS and CMS at $\sqrt{S}=$~13 TeV~\cite{CMS:2018lgn, ATLAS:2023hul}, where a combination is not yet available. 
In all cases, we consider the sum of the cross sections for $t + X$ and $\bar{t}+X$ production. In principle the ratio of cross sections for $t + X$ and $\bar{t}+X$ production can be used to constrain the $u/d$ quark ratio. 
However, we do not follow this path, considering that this ratio is already well constrained by DY data, and that the distinction between $t + X$ and $\bar{t} + X$ events is still prone to large uncertainties, also related to the treatment of fragmentation in single-top production.
The used datasets mentioned above are also listed in the website of the LHC Top Working Group.

Third, the fit includes the available datasets of   normalized cross sections double-differential in $M({t\bar{t}})$ and $y({t\bar{t}})$
at $\sqrt{S}=13$~TeV~\cite{top18004, top20001,  a190807305, a200609274}. 
This is the most detailed information on top-quark production available at the moment\footnote{A new analysis, updating Ref.~\cite{top18004}, has been recently produced by the CMS collaboration in Ref.~\cite{CMS:2024ybg}, making use of the full luminosity available in Run~2.}.

In order to estimate the range of the partonic momentum fraction $x$ probed by these data, in Fig.~\ref{fig:xparton} we show the NNLO predictions calculated with the ABMP16 PDFs for the $t\bar{t}+X$ cross sections as a function of $x$ at $\sqrt{S}=13$~TeV.
Besides the total $t\bar{t}+X$ cross section, we plot the cross sections in several kinematic regions, considering the bins from the binning scheme of the experimental measurement~\cite{top20001} in which the lowest and highest $x$ values are probed. 
From this plot, we expect these data to probe the PDFs in the range $0.002<x<0.7$.%

\subsection{Methodology/statistical analysis}
\label{sec:metho}
The methodology to perform this fit is the same as in Ref.~\cite{Alekhin:2017kpj}, using the Hessian approach to compute eigenvectors and uncertainties. A tolerance criterion $\Delta \chi^2 = 1$ is applied to the $\chi^2$, defined as in the previous ABMP16 paper. Further details can be found in Ref.~\cite{Alekhin:2017kpj}.

For the analyses of the top-quark data, covariance matrices are built according to the information on uncorrelated and correlated uncertainties reported in the experimental publications. The lack of information concerning correlations between the systematic uncertainties on the results of different analyses, not provided by the experimental collaborations, at least so far, is alleviated by the use of normalized differential cross sections which benefit from (partial) cancellation of many systematic uncertainties. Efforts towards combinations of results of different analyses, within a same experimental collaboration, or among different experimental collaborations, would be of great help towards a better understanding and control of this missing information, so far not included in our covariance matrices, with the exception of the case of total cross sections for $t\bar{t} + X$ production at $\sqrt{S}=$~7~and~8~TeV,  where an ATLAS+CMS combination has already been performed~\cite{ATLAS:2022aof} and the corresponding correlated uncertainties accounted for in our covariance matrix. 

\begin{figure}[h!]
  \begin{center}
  \includegraphics[width=0.48\textwidth]{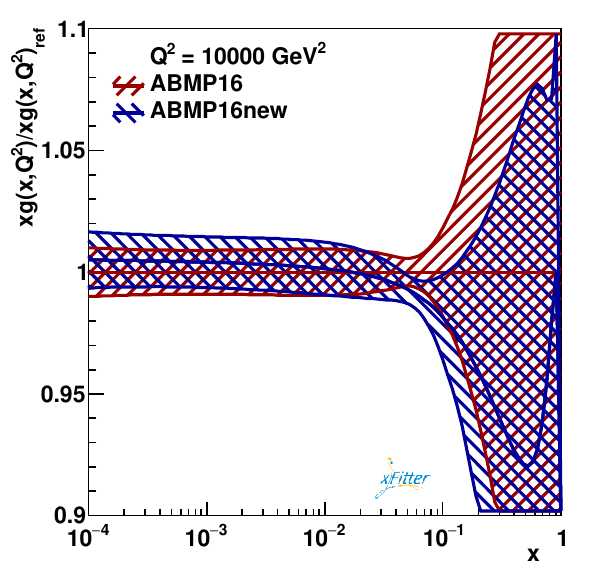}
  \includegraphics[width=0.48\textwidth]{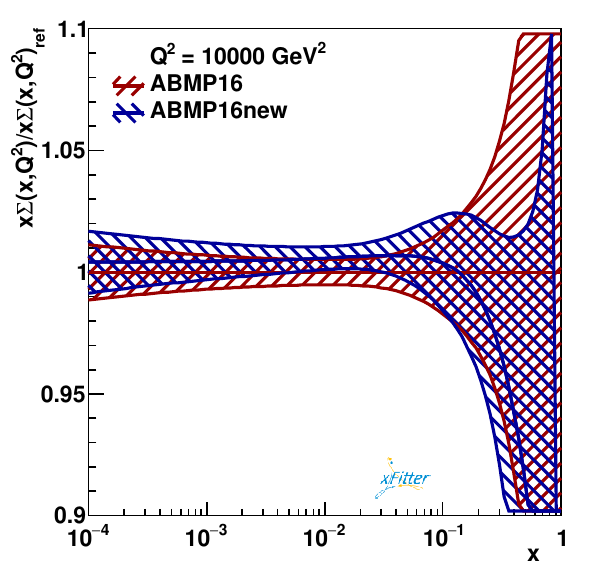}
  \includegraphics[width=0.48\textwidth]{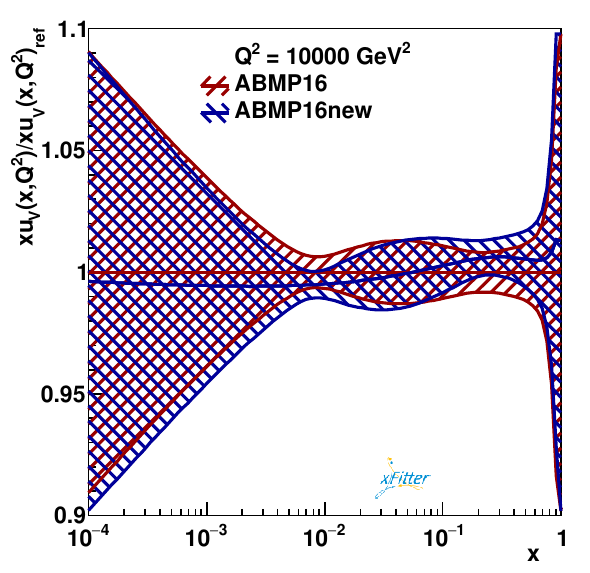}
  \includegraphics[width=0.48\textwidth]{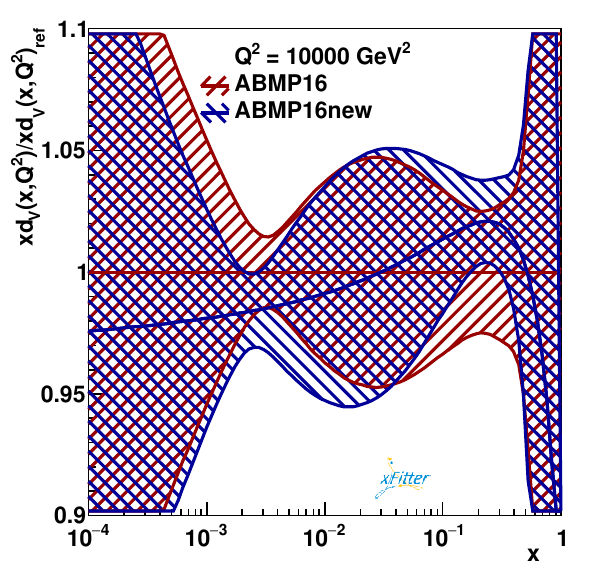}
       \caption{\label{fig:compa16vs16new_nf=5}
         Ratio of the gluon, sea-quark, valence up-quark, valence down-quark PDFs in the ABMP16new fit to the published ABMP16 fit
at a scale $Q=100$~GeV
         for $n_f=5$.
         See the text for more detail.}
       \end{center}
       \end{figure}
\begin{figure}[h!]
  \begin{center}
  \includegraphics[width=0.48\textwidth]{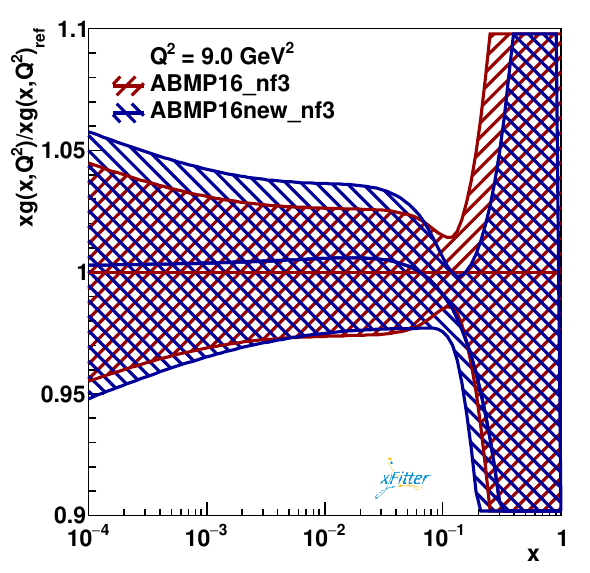}
  \includegraphics[width=0.48\textwidth]{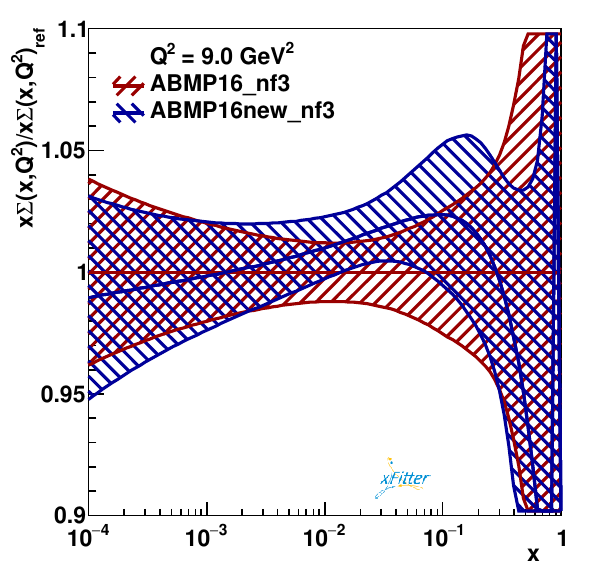}
  \includegraphics[width=0.48\textwidth]{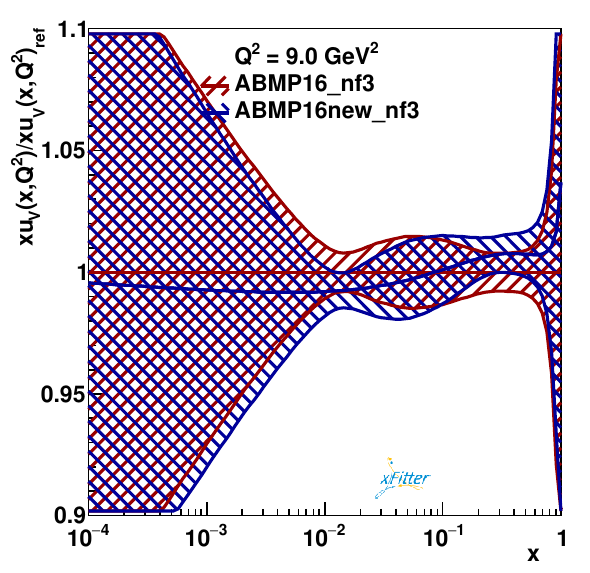}
  \includegraphics[width=0.48\textwidth]{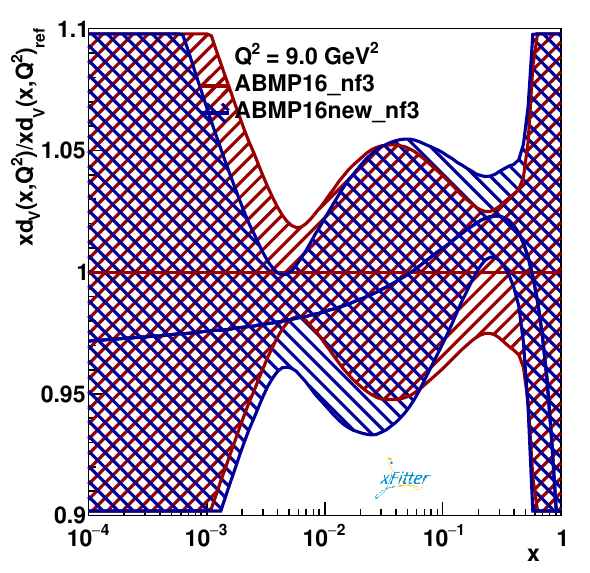}
       \caption{\label{fig:compa16vs16new_nf=3}
         Same as Fig.~\ref{fig:compa16vs16new_nf=5}, but for $n_f=3$
         and a scale of $Q=3$~GeV.}
       \end{center}
\end{figure}
In order to perform the ABMPtt fit, which includes a wide range of top-quark datasets as explained in subsection~\ref{sec:exp}, we have prepared grids for $t\bar{t} + X$ differential predictions with \texttt{MATRIX+PineAPPL}, as explained in subsection~\ref{sec:theo}, using as a basis for their computation the ABMP16 fit. 
These grids can be precomputed just once, and, being independent of the detail of the PDF parameterization, they can then be used during an iterative procedure of PDF fitting, even involving many iterations. 
Given that DY data are also included in the fit, and the computations leading to the corresponding NNLO theory predictions are quite CPU intensive, a situation by no means different from the case of top-quark production (see e.g.~Ref.~\cite{Alekhin:2021xcu}), we have made use of precomputed grids of theory predictions even for the DY case.
For consistency, we have computed even these grids by using the ABMP16 PDFs as a basis. This led even to an ``intermediate'' PDF fit, dubbed ABMP16new, that differs from the ABMP16 PDFs publicly available fit just because of this detail. 
In fact, the ABMP16 fit was performed by using as a basis in-house tables of predictions for DY cross sections computed with \texttt{FEWZ} (version 3.1)~\cite{Li:2012wna,Gavin:2012sy} from a previous PDF fit, ABM12~\cite{Alekhin:2013nda}, and then using linear interpolation on top of these tables, according to the methodology described in the Appendix of Ref.~\cite{Alekhin:2013nda}. 
Linear interpolation neglects the effects of parabolic terms. These in-house tables, differently from the grids that one can get nowadays from advanced tools like e.g. \texttt{PineAPPL}, depend indeed on the PDF parameterization, and, therefore, they need to be recomputed at each iteration of a fit, starting from a new PDF. Computing a new grid at each i\-te\-ration becomes prohibitive in terms of CPU time when performing multiple iterations, and this is the reason why, during the ABMP16 fit, only one grid was computed and used.    
The ABMP16new fit, involving a second calculation/recalculation of that grid, using as input the ABMP16 fit, instead of ABM12, is less sensitive to the missing parabolic terms and more robust than the ABMP16 fit. 
A comparison of the PDFs for gluon, sea quarks and valence quarks in the ABMP16new and ABMP16 fit for $n_f=5$ at a scale $Q = 100$~GeV
and for $n_f=3$ at a scale $Q = 9$~GeV is shown in Figs.~\ref{fig:compa16vs16new_nf=5} and~\ref{fig:compa16vs16new_nf=3}, respectively. 
It is clear from those plots that the two fits are compatible with each other within uncertainties for most of the $x$ values.

On the other hand, predictions for DIS and total cross-sections for top-quark production are computed on the fly at each iteration the same way for all ABMP fit variants, without making use of any grid or precomputed storage.

In summary, the ABMPtt fit includes DY and all other non-top-quark data as used in the ABMP16new fit, and differs from the ABMP16 fit due to a more robust procedure of inclusion of DY data, as discussed above, and because it accounts for many more top-quark datasets (see Appendix~\ref{appB:allpdfs} for comparisons between the PDF variants). The new datasets are indeed the main source of differences between the ABMPtt and ABMP16 PDFs, leading to smaller uncertainties in ABMPtt, and this will be the main focus in the following sections.

\section{Results of the $\textrm{ABMPtt}$ fit}
\label{sec:res}

We are now ready to present results of the ABMPtt fit.
Subsection~\ref{sec:pulls} is devoted to a detailed analysis of the data, checking compatibility of the datasets by performing several variants of the fit.
The results for the new PDFs are discussed in subsection~\ref{sec:pdfs}, also comparing with PDFs determined by other groups, 
and the values for $\alpha_s(M_Z)$ and the top-quark mass $m_t(m_t)$ at NNLO in the \msbar scheme are presented in subsection~\ref{sec:mtop}.
Subsection~\ref{sec:lhapdf} contains a brief description of the grids for the {{\texttt{LHAPDF}} library}~\cite{Buckley:2014ana}.

\subsection{Comparison to $t\bar{t}+X$ data}
\label{sec:pulls}

\begin{figure}[t!]
\begin{center}
  \includegraphics[width=0.8\textwidth]{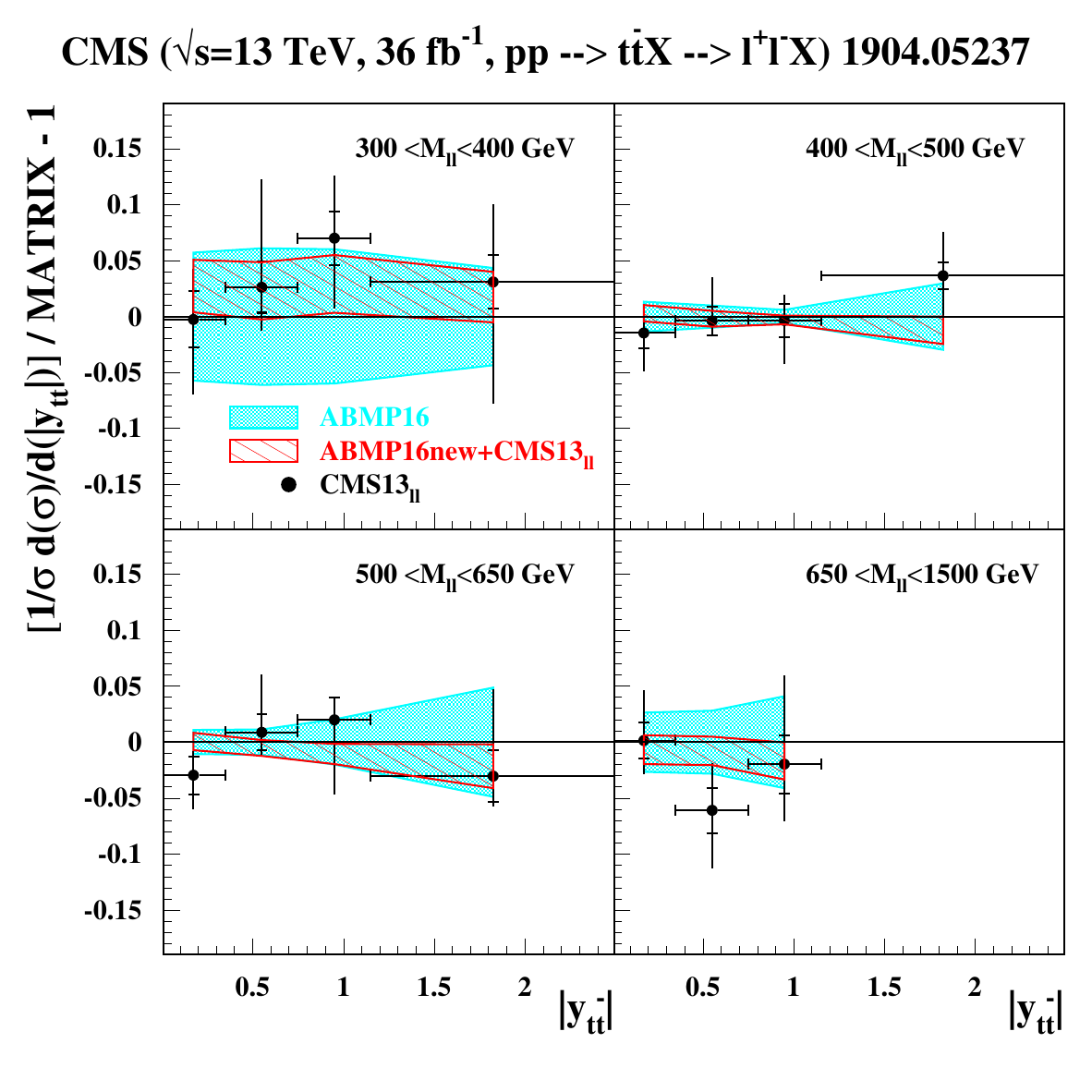}
\vspace*{-5mm}
\caption{\label{fig:cmsdile} Pulls of CMS dileptonic data from the
  analysis of Ref.~\cite{top18004}
  with respect to predictions using as input the ABMP16 fit (light-blue solid band)
  and the fit including besides all ABMP16 datasets the specific double-differential cross-section dataset indicated (red dashed band).}
\end{center}
\end{figure}
\begin{figure}[h!]
\begin{center}
  \includegraphics[width=0.8\textwidth]{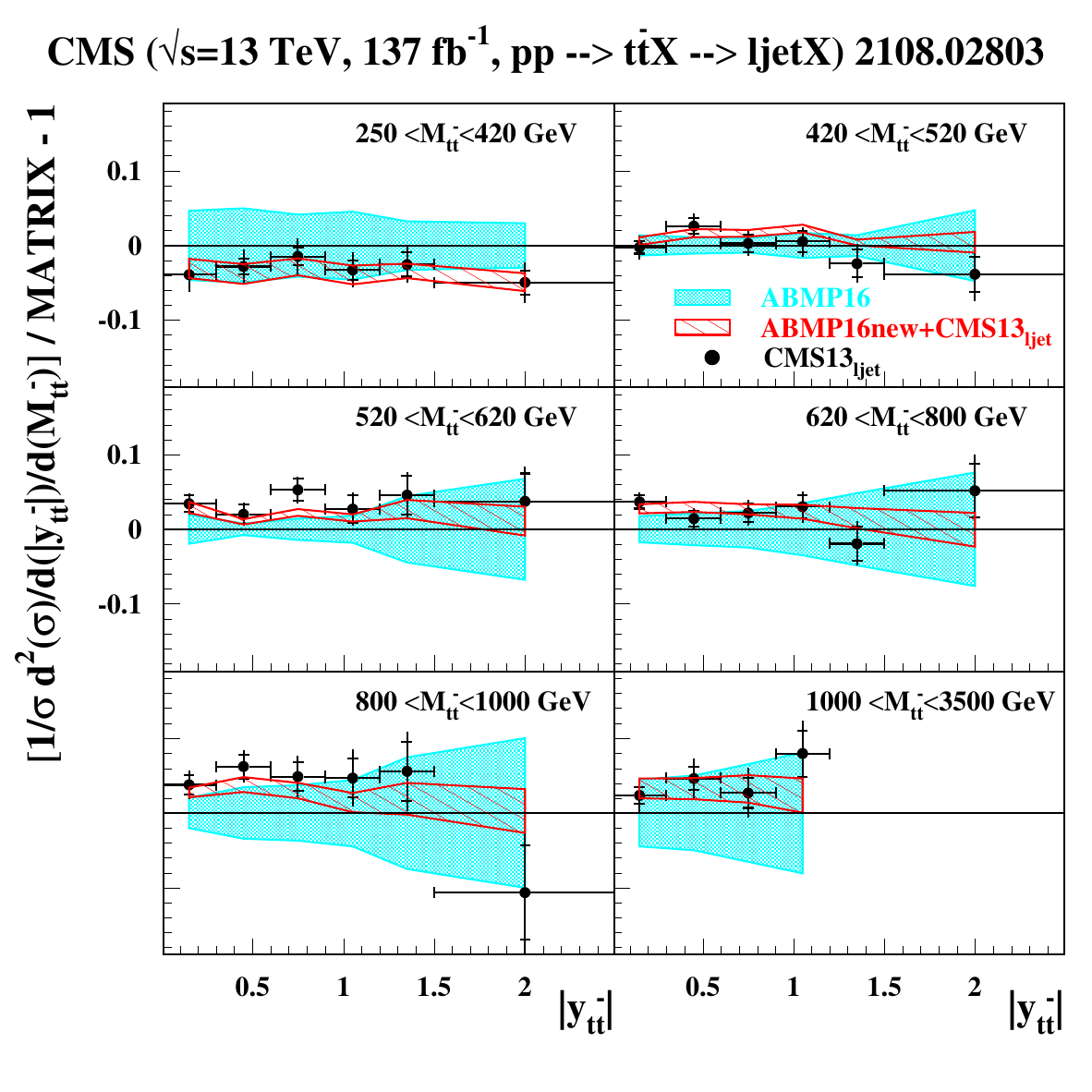}
\vspace*{-5mm}
\caption{\label{fig:cmssemilep} Same as Fig.~\ref{fig:cmsdile}, but for the CMS semileptonic data from the
  analysis of Ref.~\cite{top20001}.}
\end{center}
\end{figure}
\begin{figure}[h!]
\begin{center}
  \includegraphics[width=0.8\textwidth]{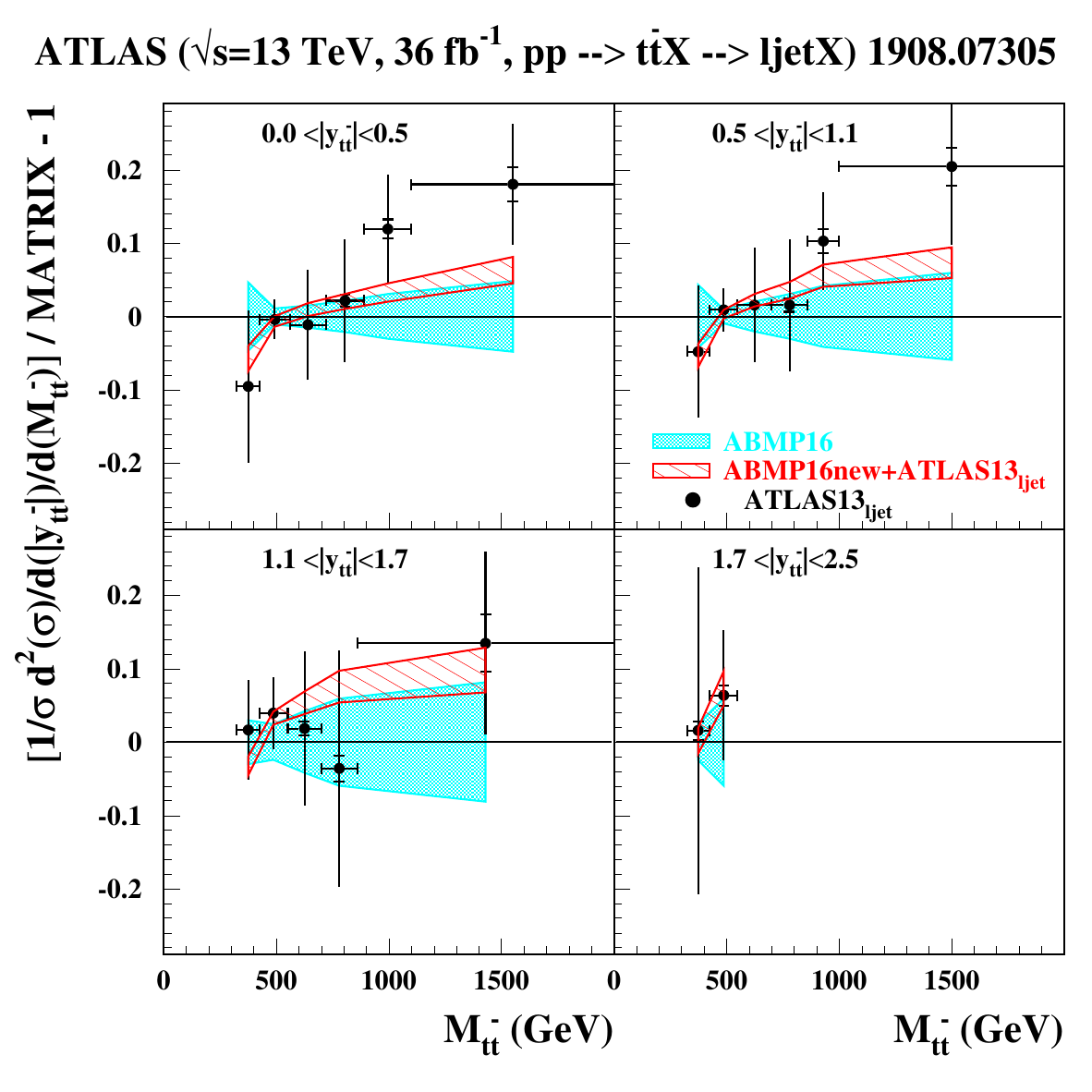}
\vspace*{-5mm}
\caption{\label{fig:atlassemilep} Same as Fig.~\ref{fig:cmsdile}, but for the ATLAS semileptonic data from the
  analysis of Ref.~\cite{a190807305}.}
\end{center}
\end{figure}
\begin{figure}[h!]
\begin{center}
\includegraphics[width=0.8\textwidth]{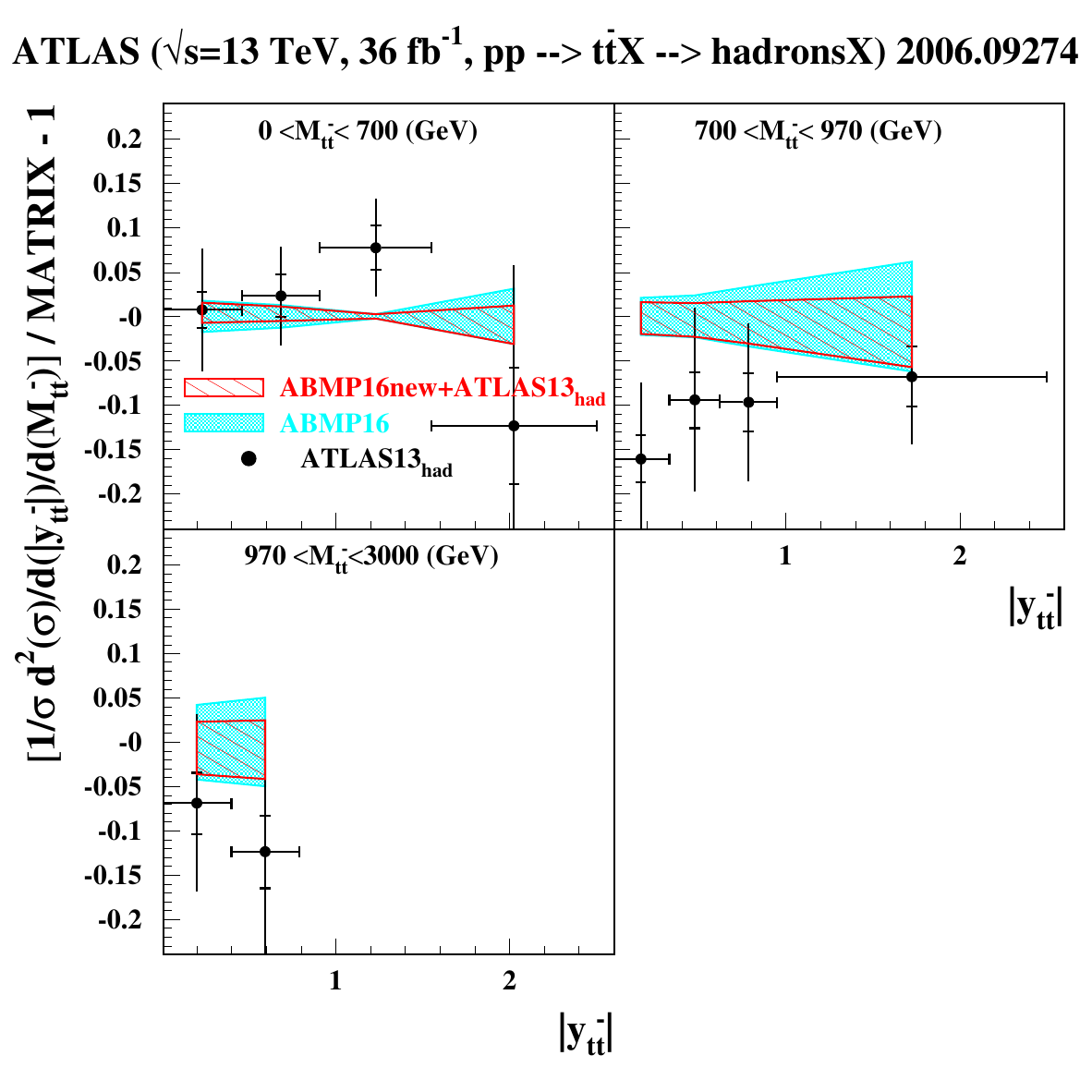}
\vspace*{-5mm}
\caption{\label{fig:atlashad} Same as Fig.~\ref{fig:cmsdile}, but for the ATLAS all-hadronic data from the
  analysis of Ref.~\cite{a200609274}.}
\end{center}
\end{figure}
The pulls of the double-differential cross-section data collected in the $t\bar{t}+X$ CMS dileptonic, CMS semileptonic, ATLAS semileptonic and ATLAS all-hadronic analyses are shown in 
Figs.~\ref{fig:cmsdile},~\ref{fig:cmssemilep},~\ref{fig:atlassemilep}, and~\ref{fig:atlashad}, 
with respect to theory predictions using as input the original ABMP16 PDF fit, and an ABMPtt fit variant obtained after including only the specific dataset 
(dubbed ``ABMP16new+specific dataset'' in the panels of the plots).
In all cases,  the uncertainty band on the pulls computed with the new fit turns out to be 
smaller than the ABMP16 one. This applies even to the all-hadronic analysis, notwithstanding the still large uncertainties that accompany it. 
We also observe that the strongest constraints are derived by the data from the CMS semileptonic analysis, corresponding to the largest integrated lu\-mi\-no\-si\-ty ($L=137$~fb$^{-1}$), whereas the worst agreement data/theory is obtained in case of the ATLAS semileptonic analysis, where the data tend to be systematically larger than theory predictions at large $M({t\bar{t}}) \sim 1500$~GeV in all $|y({t\bar{t}})|$ bins. This tendency is not observed in the pulls of the CMS semileptonic analysis, characterized by a larger $L$ and is also opposite with respect to the trend    shown at large $M({t\bar{t}})$ in the ATLAS all-hadronic analysis, characterized however
by larger uncertainties. 
An ATLAS semileptonic analysis with better statistics  
would certainly be of help for understanding the reasons of these differences and solve the discrepancy.

\begin{table}[h!]
\begin{center}                   
\begin{tabular}{|c|c|c|c|}
    \hline
Dataset & $NDP$ & $\chi^2$ & 
$m_t(m_t)$ (GeV)
\\
\hline
ATLAS$13_{ljet}$  & 19 & 25.2 & $158.9\pm1.3$ \\
ATLAS$13_{had}$  & 10 & 11.3 &  $160.5\pm 2.0$ \\
CMS$13_{ll}$ & 15 & 13.9 &   $161.1\pm1.4$ \\
CMS$13_{ljet}$ & 34 & 37.4 & $158.7\pm0.9$ \\
\hline
 \end{tabular}
\caption{\label{tab:tab3modified}
  {The values of {$\chi^2$} and the top-quark mass {$m_t(m_t)$} for each single 
    {$t \bar{t} +X $} dataset obtained in the
    variants of present analysis including just the specific
    {$t \bar{t} + X$} double-differential cross-section
    dataset.}
}
\end{center}
\end{table}

The $\chi^2$/dataset corresponding to the scenarios of Figs.~\ref{fig:cmsdile},~\ref{fig:cmssemilep},~\ref{fig:atlassemilep}, and~\ref{fig:atlashad} are reported in Table~\ref{tab:tab3modified}. 
We observe reasonable values of $\chi^2$, as compared to the number of data points (NDP), with a slightly worse agreement of data/theory for the ATLAS semileptonic analysis, in line with the discussion above. 
In the same table, we also report the value of the top-quark mass renormalized in the \msbar scheme, fitted simultaneously to the PDFs
and $\alpha_s(M_Z)$ in each case. The extracted values of $m_t(m_t)$ are compatible among each other, considering the $1\sigma$ uncertainty on each of them. 

\begin{table}[h!]
\begin{center}                   
  \begin{tabular}{|l|c|c|c|c|c|c|c|}
    \hline
Experiment & Dataset &
{$\sqrt{s}$ (TeV)} & {$NDP$}& 
\multicolumn{3}{c|}{{$\chi^2$}}
\\
\hline
 \multicolumn{4}{|c|}{} &{I}&{II}& {III} \\ 
\hline
ATLAS & ATLAS$13_{ljet}$&13 & 19 & 34.0 (42.8)& 28.2&--
\\
 & ATLAS$13_{had}$&13 & 10 & 11.9 (11.7)& 11.6&--
\\
\hline
CMS & CMS$13_{ll}$&13 & 15 & 20.7 (15.9)& --&19.6
\\
 & CMS$13_{ljet}$&13 & 34 & 44.3 (52.0)& --&42.4
\\
\hline
 \end{tabular}
\caption{\label{tab:chi2modified}
The values of {$\chi^2$} obtained for various 
{$t \bar{t} + X$} datasets included in different variants of the present analysis
(column {I}: variant including both ATLAS and CMS double-differential cross-section datasets; 
column {II}: variant including only ATLAS ones; 
column {III}: variant including only CMS ones).
The figures in parenthesis give the values of $\chi^2$ obtained for the corresponding dataset using as input the central ABMP16 PDFs without re-fit. 
}
\end{center}
\end{table}

After the variants of analyses including a single $t\bar{t} +X$ dataset at a time, presented above, we consider other variants, as detailed  in Table~\ref{tab:chi2modified}. 
In particular, in the analyses where the two ATLAS datasets are included (column II) or the two CMS datasets are included (column  III), the $\chi^2$s
slightly worsen with respect to the cases of including only a single dataset at a time, as reported in Table~\ref{tab:tab3modified}. 
Overall, the $\chi^2$s values in Table~\ref{tab:chi2modified} 
are still mostly compatible with the previous ones in Table~\ref{tab:tab3modified}, 
except for the CMS dileptonic dataset, where the $\chi^2$ value rises from 13.9 to 19.6. 
This points to a slight tension with the CMS semileptonic analysis, 
which seems also confirmed by the fact that the best description of the CMS dileptonic data turns out to be given by the original ABMP16 fit, which, however, provides a less accurate  description of the CMS semileptonic data (see values in parentheses in column~I of Table~\ref{tab:chi2modified}).  

The $\chi^2$s of the analysis including all CMS and ATLAS datasets (column I of Table~\ref{tab:chi2modified}) are slightly worse than the ones of the analysis including just the CMS datasets (column III), but are still of  comparable quality.
A comparison of the fit including both ATLAS datasets (column II) with the one of column I in Table~\ref{tab:chi2modified} shows that the $\chi^2$ value for the all-hadronic dataset slightly worsens when including also the datasets from CMS, but it is still compatible with the previous one.
On the other hand, the $\chi^2$ value for the ATLAS semileptonic dataset definitely worsens when considering both ATLAS and CMS datasets (column I of Table~\ref{tab:chi2modified}, compared to column II), which displays the tension of the ATLAS semileptonic dataset with all other ones. This particular dataset is also not described well by the original central ABMP16 PDFs (see the value in parenthesis in column I).

In all variants of ABMPtt analysis discussed in Table~\ref{tab:chi2modified}, besides selected double-differential datasets, recent  datasets for $t\bar{t}+X$ and the sum of $t+X$ and $\bar{t}+X$ total inclusive cross sections, as detailed in Section~\ref{sec:exp}, updated with respect to the case of the ABMP16 fit, have also been included.
The single-top data ($t + X$ and $\bar{t} + X$) are not yet at the same level of precision as the $t\bar{t} + X$ ones. 
However, we have included them for checking the consistency and we observe that they improve the fit stability. 
Additionally, they help in reducing the correlations between the top-quark mass $m_t(m_t)$ and the strong coupling $\alpha_s$, which emerges from the analysis of $t\bar{t} + X$ data.
The reason is that single-top hadroproduction in the $s$- and $t$-channels 
is an 
electroweak
process at
leading order, thus reducing the sensitivity to $\alpha_s$.
In the past, single-top data were also used standalone to constrain $m_t(m_t)$~\cite{Alekhin:2016jjz} using different PDFs as input.
This has led to values for $m_t(m_t)$ compatible with our present fit result, although characterized by a larger uncertainty ($1 \sigma \sim$~3~GeV). 

\begin{figure}[h!]
  \includegraphics[width=0.9\textwidth]{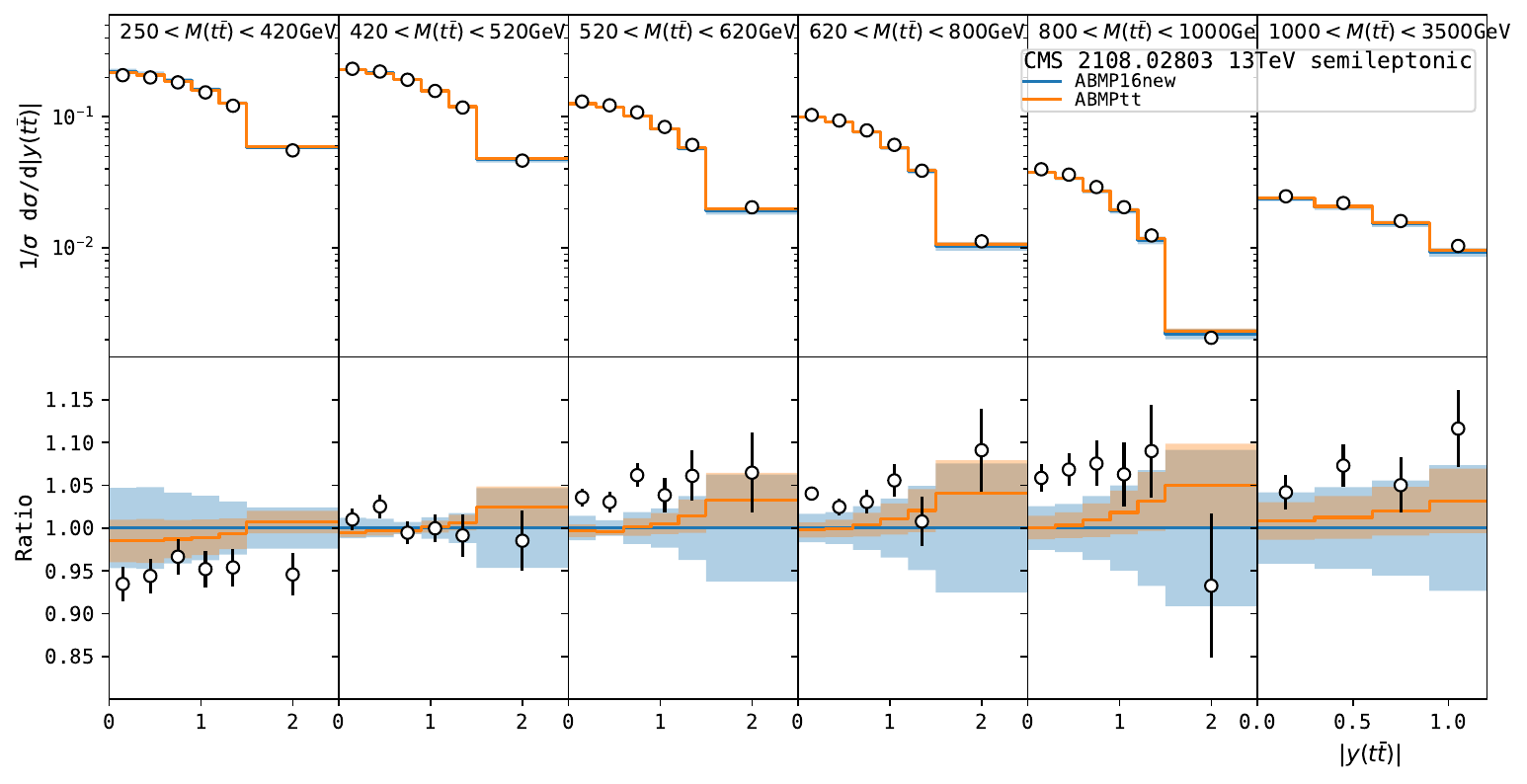}
  \caption{\label{fig:compdiffe} 
  NNLO theory predictions using as input the ABMPtt and the ABMP16new PDFs for $|y(t\bar{t})|$ distributions in different
    $M({t\bar{t}})$    bins vs. CMS experimental data at
$\sqrt{S}=13$~TeV    from Ref.~\cite{top20001}.}
\end{figure} 
\begin{figure}[h!]
  \includegraphics[width=0.9\textwidth]{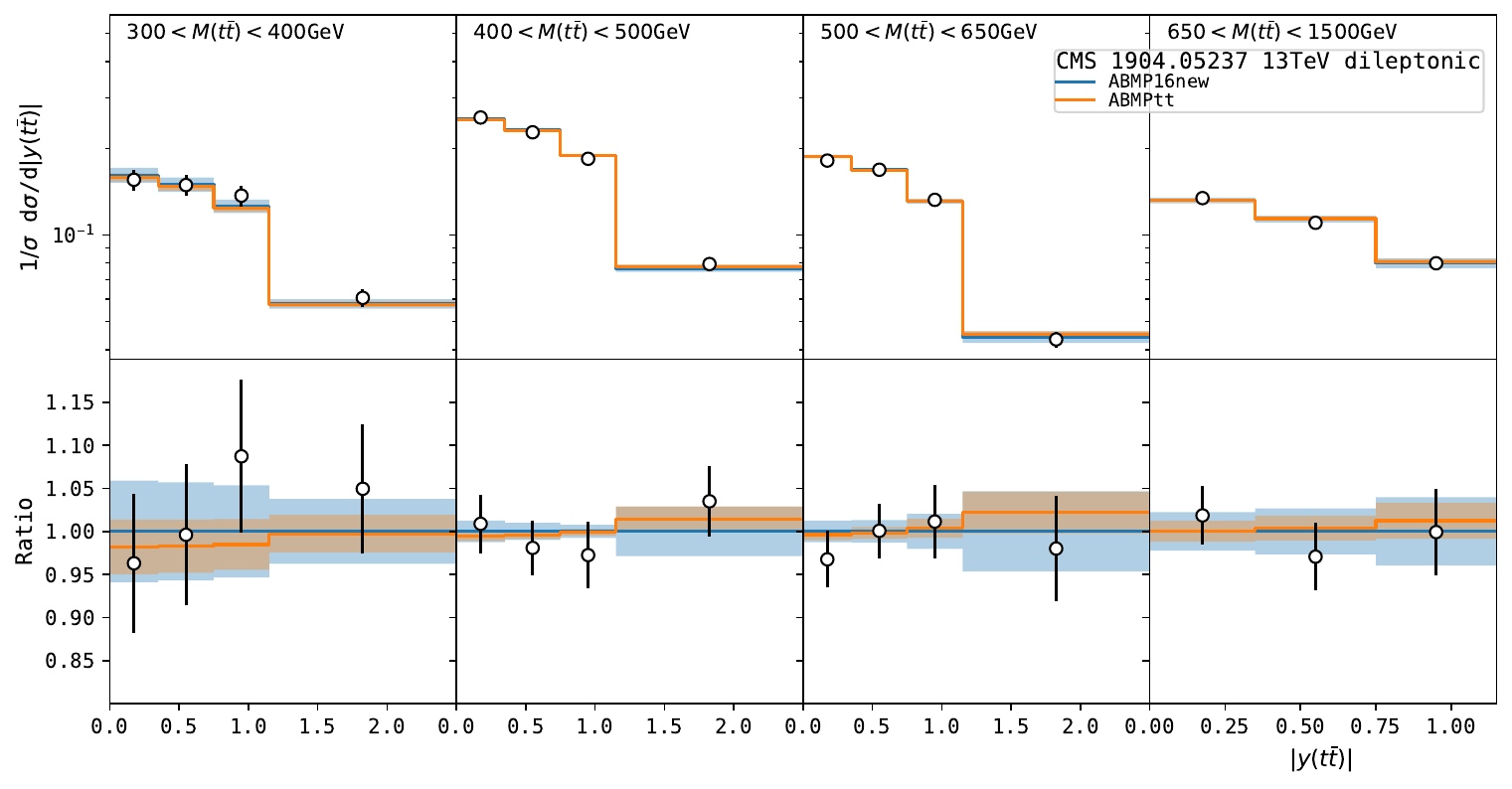}
  \caption{\label{fig:compdiffe18}
    NNLO theory predictions using as input the ABMPtt and ABMP16new PDFs for
$|y(t\bar{t})|$
    distributions in different     $M({t\bar{t}})$ bins  vs. CMS experimental data
    at $\sqrt{S}=13$~TeV from Ref.~\cite{top18004}.}
\end{figure}
\begin{figure}[h!]
  \includegraphics[width=0.9\textwidth]{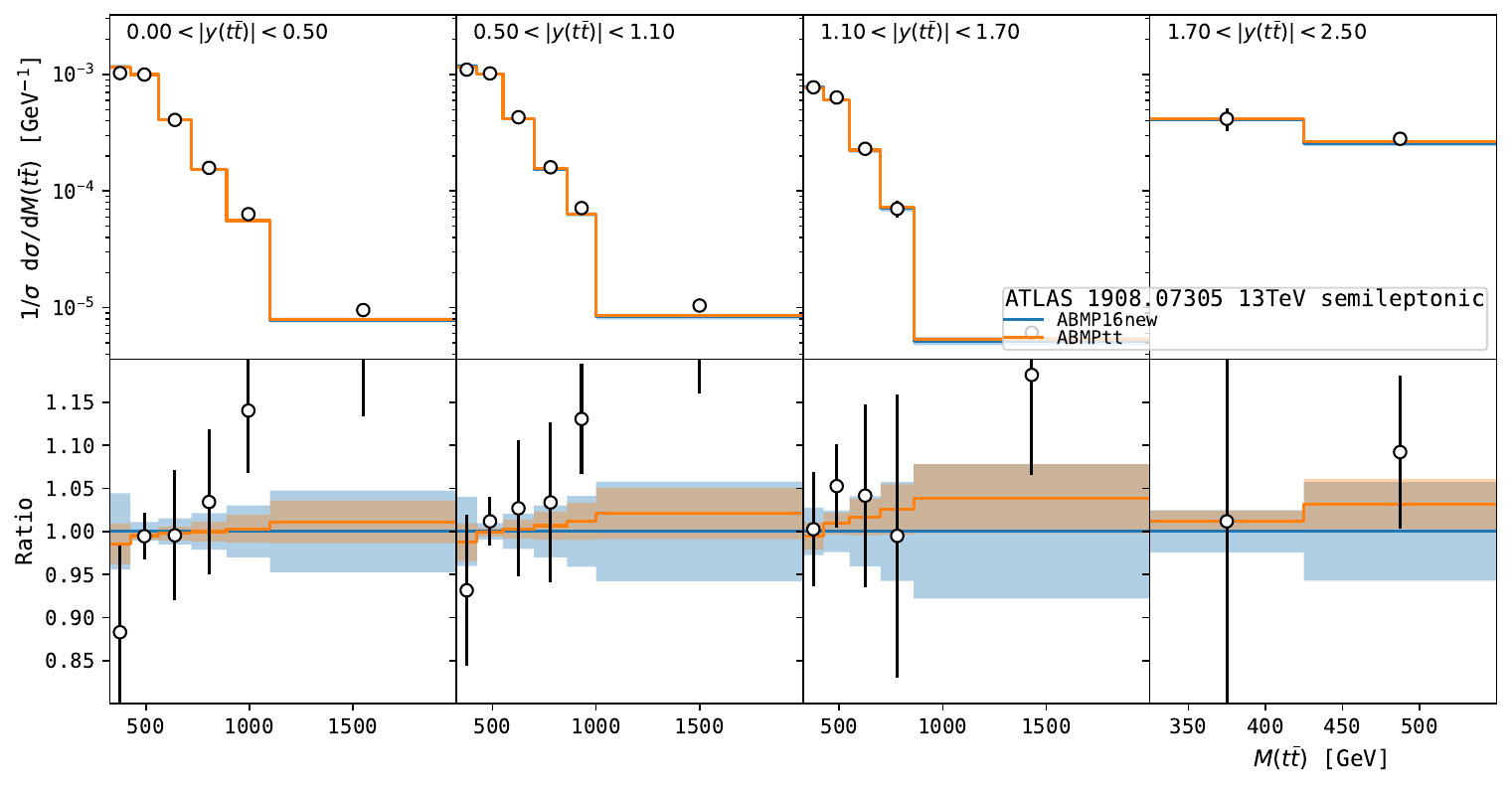}
  \caption{\label{fig:compdiffeatlas1} 
  NNLO theory predictions using as input the ABMPtt and the ABMP16new PDFs for $M({t\bar{t}})$  distributions in different
     $|y(t\bar{t})|$   bins vs. ATLAS experimental data at
$\sqrt{S}=13$~TeV    from Ref.~\cite{a190807305}.}
\end{figure} 
\begin{figure}[h!]
  \includegraphics[width=0.9\textwidth]{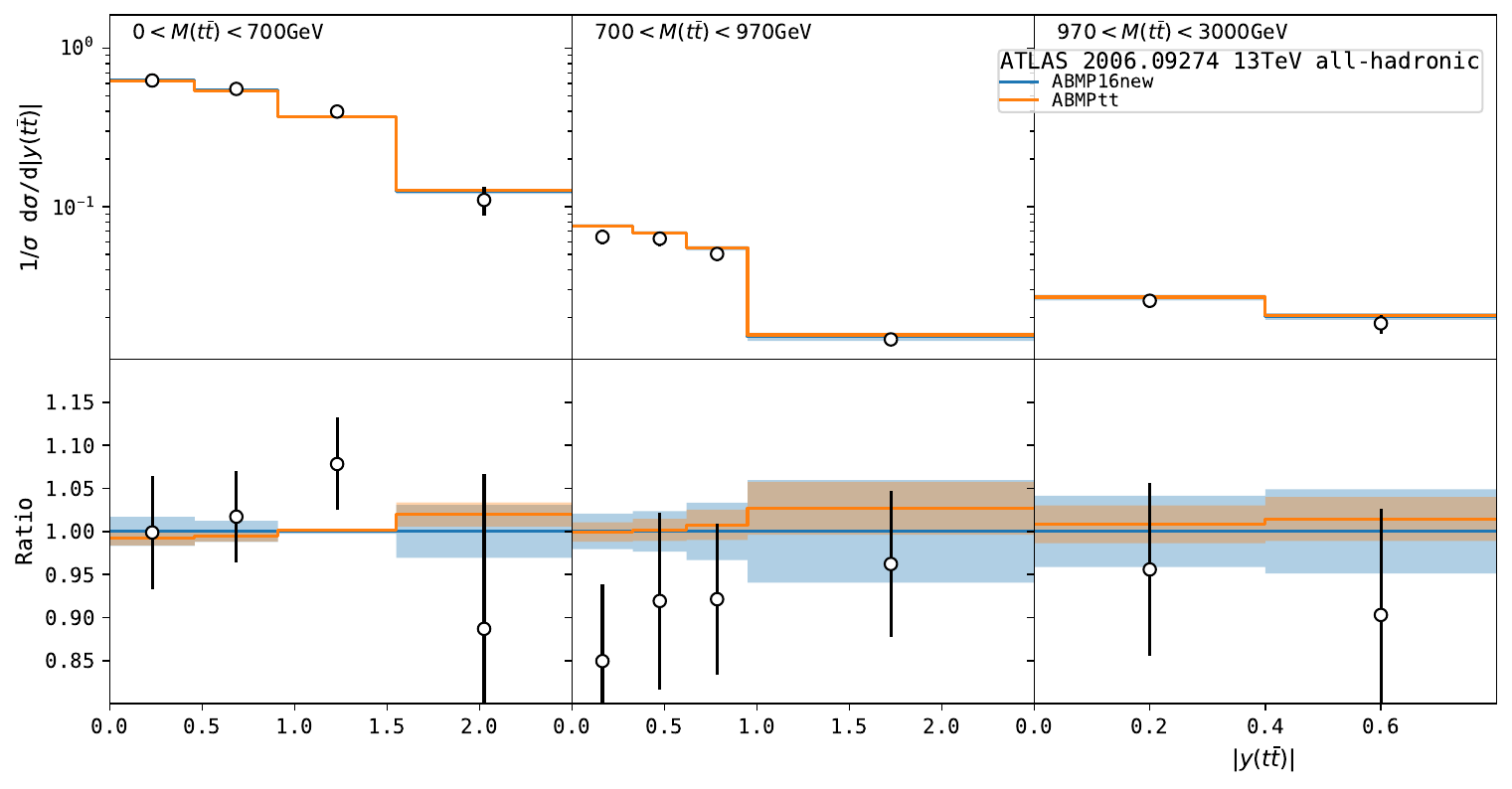}
  \caption{\label{fig:compdiffeatlas2} 
  NNLO theory predictions using as input the ABMPtt and the ABMP16new PDFs for $|y(t\bar{t})|$ distributions in different
    $M({t\bar{t}})$    bins vs. ATLAS experimental data at
$\sqrt{S}=13$~TeV    from Ref.~\cite{a200609274}.}
\end{figure} 
\begin{figure}[h!]
  \includegraphics[width=0.9\textwidth]{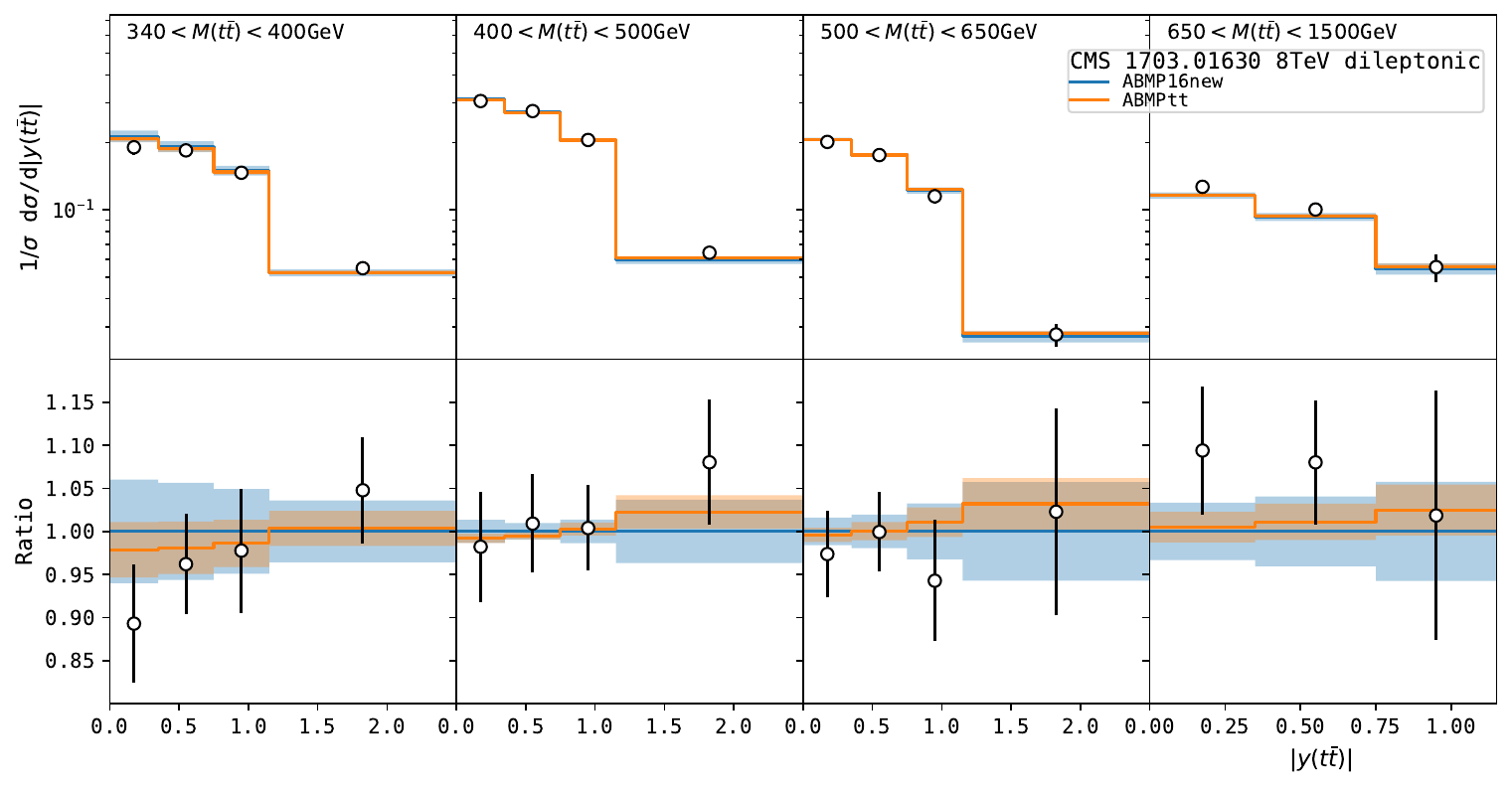}
  \caption{\label{fig:compdiffe8}
    NNLO theory predictions using as input the ABMPtt and ABMP16new PDFs for $|y(t\bar{t})|$ distributions in different $M({t\bar{t}})$ bins  vs. CMS    experimental data at $\sqrt{S} =$~8~TeV from Ref.~\cite{top14013}.}
\end{figure}
\begin{figure}
  \includegraphics[width=0.48\textwidth]{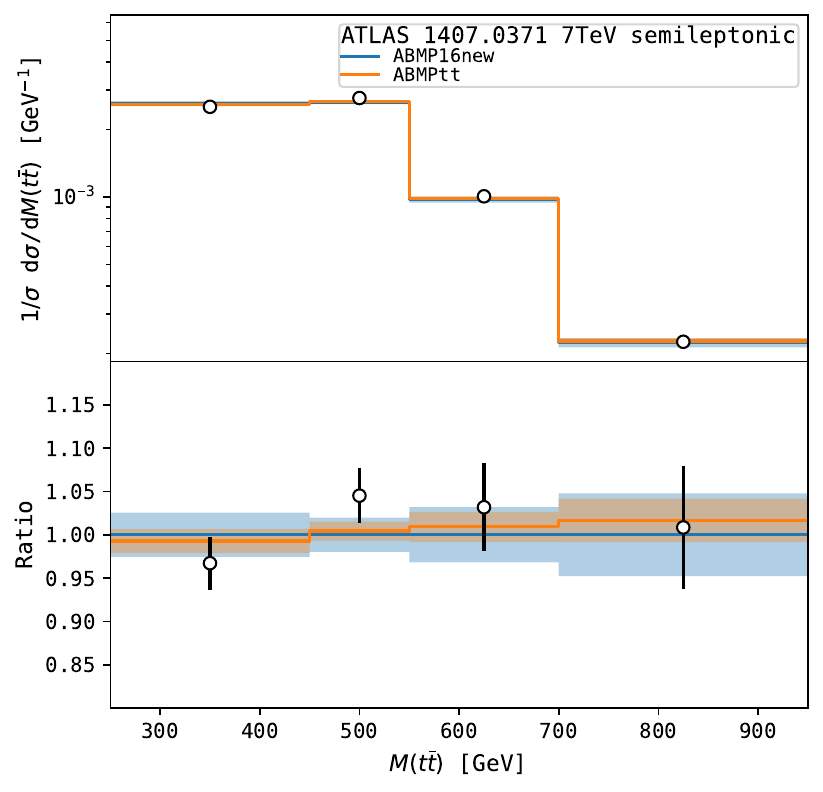}
  \includegraphics[width=0.48\textwidth]{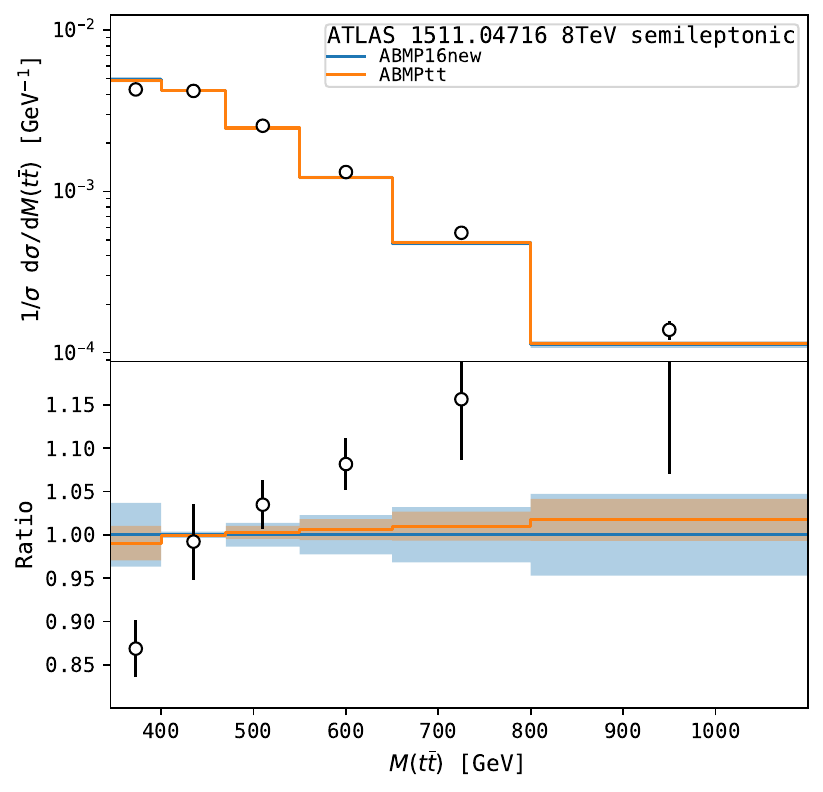}\\
  \includegraphics[width=0.48\textwidth]{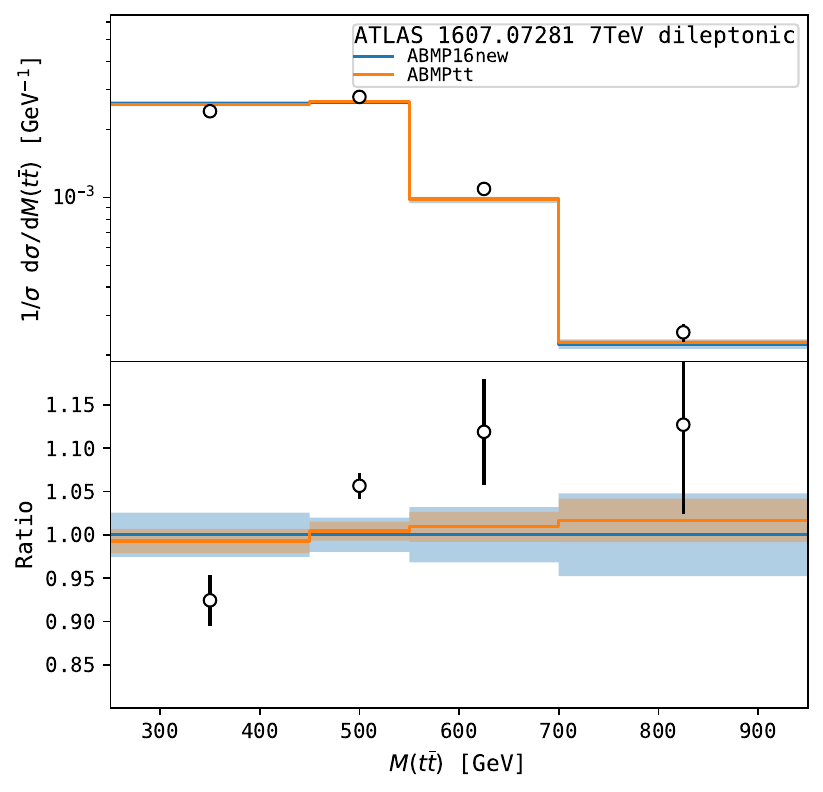}
  \includegraphics[width=0.48\textwidth]{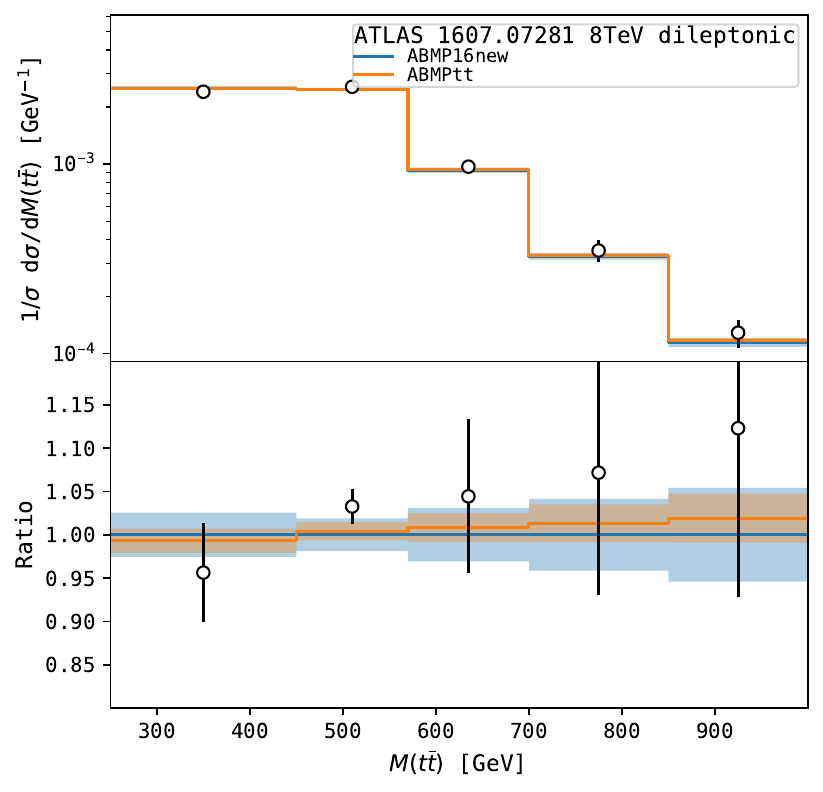}
  \caption{\label{fig:compdiffeatlas3}
    NNLO theory predictions using as input the ABMPtt and ABMP16new PDFs for $M({t\bar{t}})$ distributions  vs. ATLAS    experimental data at $\sqrt{S} =$~7~and~8~TeV from Ref.~\cite{a14070371, a151104716, a160707281}.}
\end{figure}
\begin{figure}
  \includegraphics[width=0.9\textwidth]{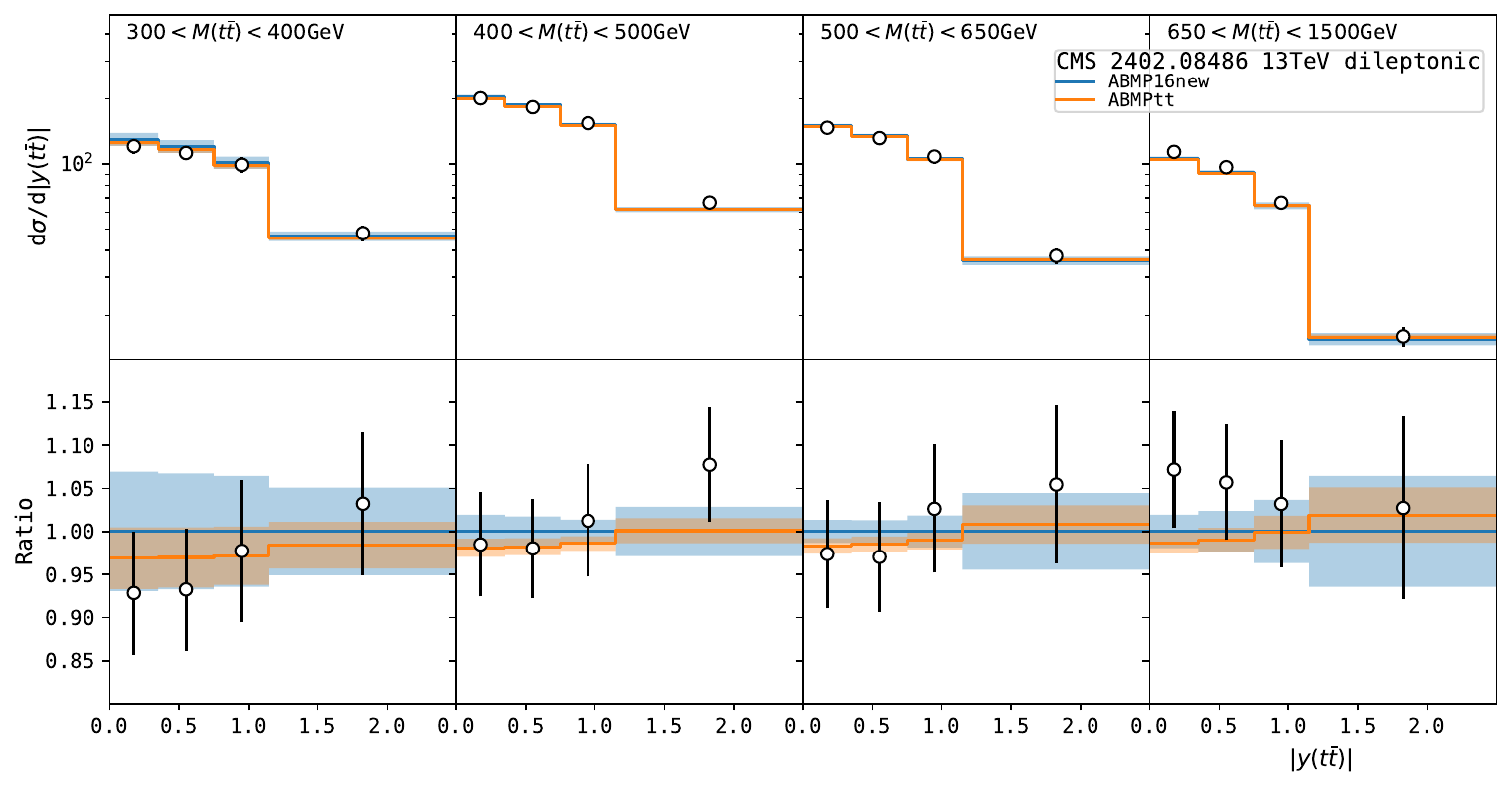}
  \caption{\label{fig:complatest}
    NNLO theory predictions using as input the ABMPtt and ABMP16new PDFs for
    $|y(t\bar{t})|$ distributions in different $M({t\bar{t}})$ bins  vs. recent CMS   experimental data at $\sqrt{S}=13$~TeV    from Ref.~\cite{CMS:2024ybg}}
\end{figure}

Having established mutual compatibility of all top-quark datasets (with the qualifications just discussed), we show in Figs.~\ref{fig:compdiffe},~\ref{fig:compdiffe18},~\ref{fig:compdiffe8},~\ref{fig:compdiffeatlas1},~\ref{fig:compdiffeatlas2},~\ref{fig:compdiffeatlas3}, and~\ref{fig:complatest} a comparison of the theory predictions at NNLO in QCD using the final ABMPtt PDFs and those based on the ABMP16new ones, as well as ratios with respect to ABMP16new central predictions, for all double-differential $t\bar{t}$ data used in the ABMPtt fit (see subsection~\ref{sec:exp}). 
The double-differential normalized $t\bar{t}+X$ cross-section CMS data in the semileptonic channel and in the dileptonic channel used for the fit are shown in Figs.~\ref{fig:compdiffe} and~\ref{fig:compdiffe18}, respectively, and we observe good compatibility of those data with the NNLO predictions of the ABMPtt PDFs.
The comparison with the double-differential normalized ATLAS data in the semileptonic and all-hadronic channel also used in the fit is shown in Figs.~\ref{fig:compdiffeatlas1} and~\ref{fig:compdiffeatlas2}. 
In all cases, the predictions with the ABMP16new PDFs, not including these data but just the same data as the ABMP16 fit and improved DY grids with respect to the latter (see subsection~\ref{sec:metho}), show also overall good agreement with the data, as could be expected from the results of Ref.~\cite{Garzelli:2023rvx}.

The $\sqrt{S}=8$~TeV double-differential distributions by CMS in the dileptonic channel from Ref.~\cite{top14013}, not used in any variant of the ABMP16 fit, but used in many other global PDF fits, are shown in Fig.~\ref{fig:compdiffe8}, 
finding agreement within theoretical and experimental uncertainties in almost all bins.
The single-differential distributions at $\sqrt{S}=7$~TeV and 8~TeV by ATLAS from Refs.~\cite{a14070371, a151104716, a160707281}, not used in any variant of the ABMP16 fit, but used in various global PDF fits, are shown in Fig.~\ref{fig:compdiffeatlas3}. 
One should observe that 
in the cases of the ATLAS semileptonic analysis at $\sqrt{S}$ = 8 TeV and of the dileptonic analysis at $\sqrt{S}$ = 7 TeV (shown as second and third panels of the figure)
theory predictions turn out to 
overestimate
data at small $M(t\bar{t})$ and 
underestimate
them at large $M(t\bar{t})$, a trend absent in the case of the ATLAS semileptonic analysis at $\sqrt{S}$~=~7~TeV
(shown \textcolor{blue}{as first panel} in the same figure), but similar to the one also observed in the analysis of ATLAS semileptonic data at $\sqrt{S}$=~13~TeV (see Fig.~\ref{fig:atlassemilep} and discussion in Section~\ref{sec:res}), although in an attenuated form. 

Finally, the comparison of NNLO theory predictions using as input the most general ABMPtt fit, as well as the ABMP16new fit, with the most recent experimental data on    inclusive double-differential cross-sections in the dileptonic channel, as published by CMS in Ref.~\cite{CMS:2024ybg}, is shown in Fig.~\ref{fig:complatest}.
As discussed in Section~\ref{sec:exp},  
this dataset is not yet included in any variant of the ABMPtt fit. One can observe a good level of agreement between theory predictions and experimental data in almost all bins.

\subsection{PDF results}
\label{sec:pdfs}

\begin{figure}[h!]
  \begin{center}
  \includegraphics[width=0.48\textwidth]{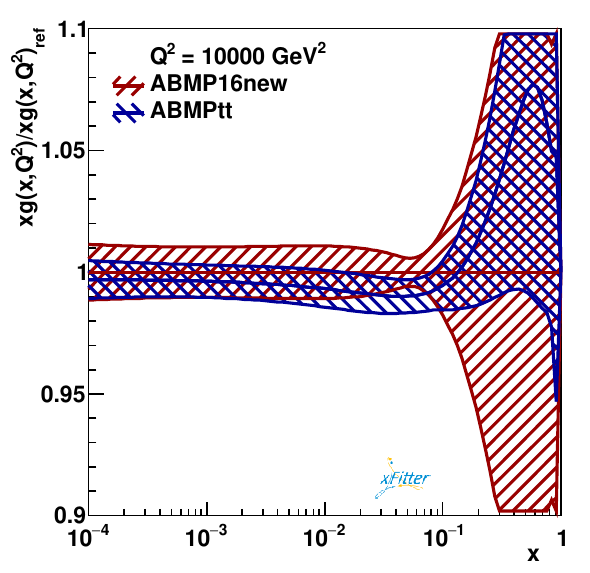}
  \includegraphics[width=0.48\textwidth]{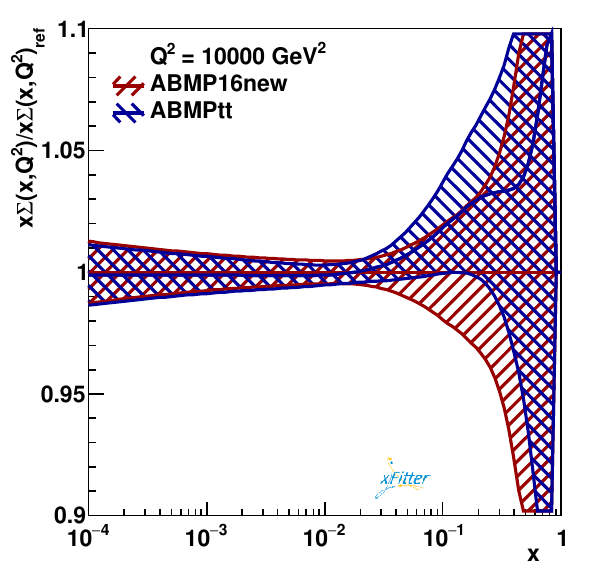}
  \includegraphics[width=0.48\textwidth]{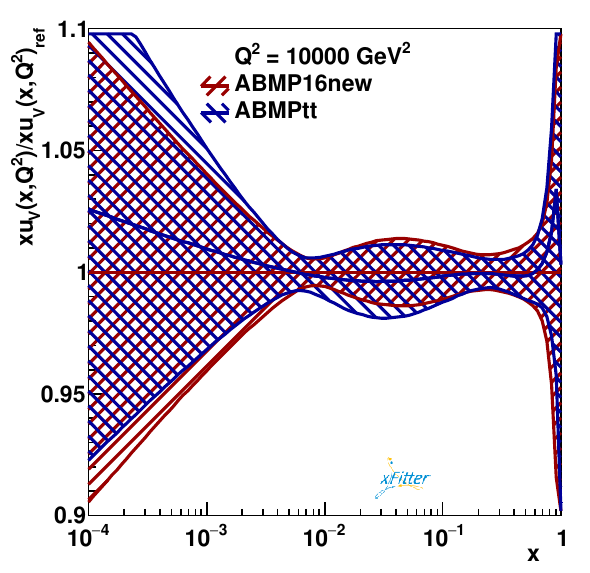}
  \includegraphics[width=0.48\textwidth]{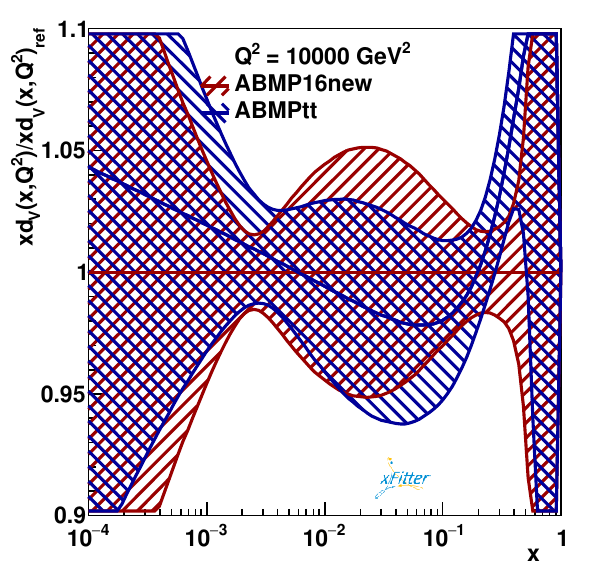}
       \caption{\label{fig:compa16ttvs16new_nf=5}
         Ratio of the gluon, sea-quark, valence up- and down-quark PDFs in the  ABMPtt fit to the variant ABMP16new at a scale $Q=100$~GeV for $n_f=5$. See the text for more detail.}
       \end{center}
\end{figure}

A comparison of the PDFs for gluon, sea and valence quarks in the ABMPtt and ABMP16new fit for $n_f=5$ and a scale $Q=100$~GeV is shown in Fig.~\ref{fig:compa16ttvs16new_nf=5}.
The gluon in the ABMPtt fit is slightly smaller (maximum 2\%) than in the baseline ABMP16new fit for $x < 10^{-1}$, whereas it becomes larger than the baseline at $x \gtrsim 0.2$, with a maximum difference of $\sim 8\%$ at $x \sim 0.5$, where, however, the uncertainties on the ABMP16new PDF fit are much larger.
Thus, the top-quark data helps to further constrain the gluon PDF and to improve its accuracy in the ABMPtt fit.
The sea-quark distribution remains approximately the same in the two fits for $x < 2 \cdot 10^{-2}$, whereas, for larger $x$ values, the sea-quark distribution in the ABMPtt fit is larger than in the baseline.
The light-flavor distribution is only moderately affected by the inclusion of the top-quark data in the fit.
For the valence up-quark PDF we observe up to 2\% difference, whereas the valence down-quark distribution shows bigger differences with respect to the baseline, especially in the region $0.2 < x < 0.5$, with the uncertainties remaining approximately the same, in both cases.
All shifts of the central values of the gluon, sea- and valence-quark distributions observed in the ABMPtt 
compared to the ABMP16new PDFs, and also the published ABMP16 PDFs, are within the $1\sigma$ uncertainty bands.
For the latter, this is shown in Appendix \ref{appB:allpdfs}.
\begin{figure}[h!]
  \includegraphics[width=0.49\textwidth]{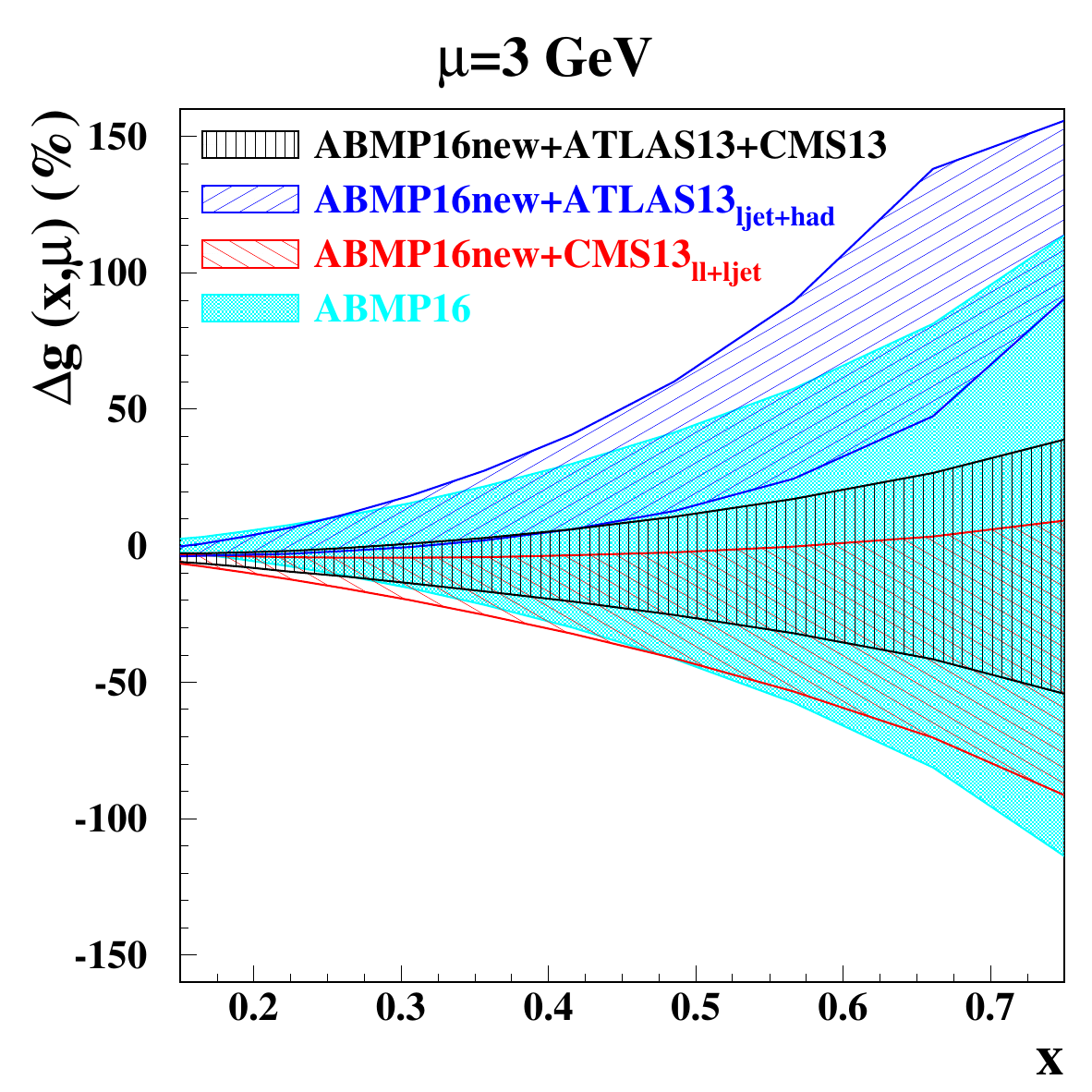}
  \includegraphics[width=0.49\textwidth]{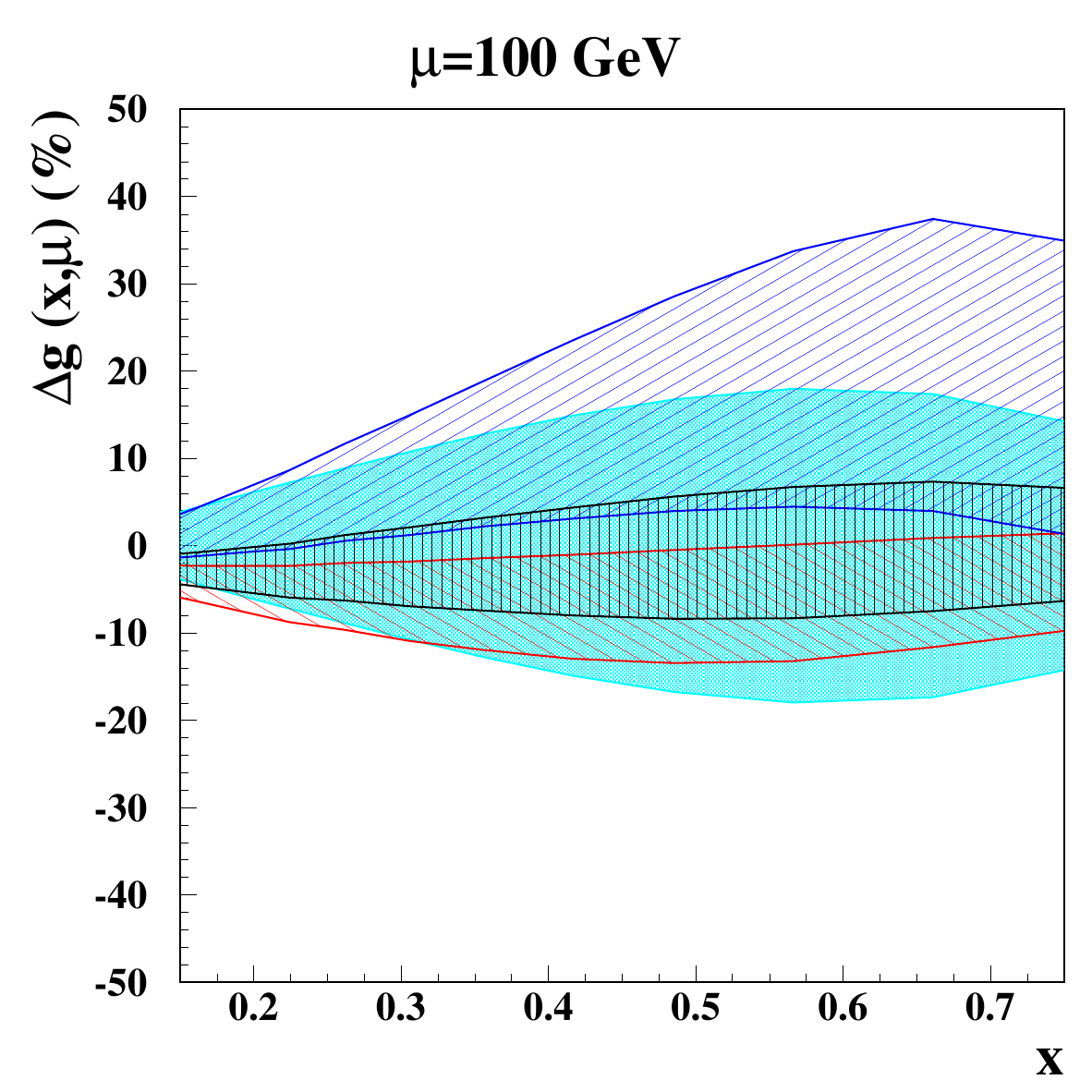}
 \caption{\label{fig:gluon} Percentage difference of the
   gluon distribution as a function of $x$ for the variants of the ABMPtt fit  of Table~\ref{tab:chi2modified}, including ATLAS
   $t\bar{t}+X$ double-differential cross-section datasets (right-tilted hatched area),
   CMS ones (left-tilted hatched areas), or both (vertical hatched area)
                     and  the ABMP16 case used as reference
                     (light-blue solid area). The left panel refers to factorization scale
                     $\mu   =$~3~GeV and $n_f=3$, whereas the right panel refers to $\mu  =$~100~GeV and $n_f=5$. }
\end{figure}

In order to illustrate in more detail the impact of top-quark data on the gluon PDF, we have extracted it from variants of the fit, where we include specific groups of datasets at a given time. In particular, in Fig.~\ref{fig:gluon} we plot the percentage differences between the gluon distributions extracted from the variants analyzed in Table~\ref{tab:chi2modified} and the ABMP16 one as a function of $x$ at the starting scale $\mu=3$~GeV and evolved to $\mu=100$~GeV.
The gluon distributions in the ABMPtt fit variants are compatible with the ABMP16 one, and present uncertainties decreased by a factor $\sim 2$ with respect to ABMP16 in the large $x$ region, $x \gtrsim 0.1$. 
The ATLAS and CMS datasets point towards opposite trends of $g(x)$ at large $x$. 
ATLAS prefers a larger $g(x)$, related to the fact that the ATLAS semileptonic data tend to be larger than theory predictions at large $M({t\bar{t}})$ (see discussion
in section~\ref{sec:pulls} and Fig.~\ref{fig:atlassemilep}).
Note that this trend is not visible for the ATLAS all-hadronic data of Fig.~\ref{fig:atlashad}, that, however, have a smaller weight in the analysis, due to the larger uncertainties.
At low scales $\mu$, the behavior of $g(x)$ in the fit including both ATLAS and CMS datasets
is dominated by the CMS semileptonic differential data, as is clear from comparing to the case where only CMS datasets are included, also shown in the left panel of Fig.~\ref{fig:gluon}. 
This trend is indeed attenuated in the right panel of 
Fig.~\ref{fig:gluon}, due to the effect of QCD evolution that couples the quark 
and gluon PDFs, reducing the differences between different variants of the fit, and makes the gluon at large $x$ also more sensitive to the value of $\alpha_s(M_Z)$. 

\begin{figure}[h!]
  \includegraphics[width=0.48\textwidth]{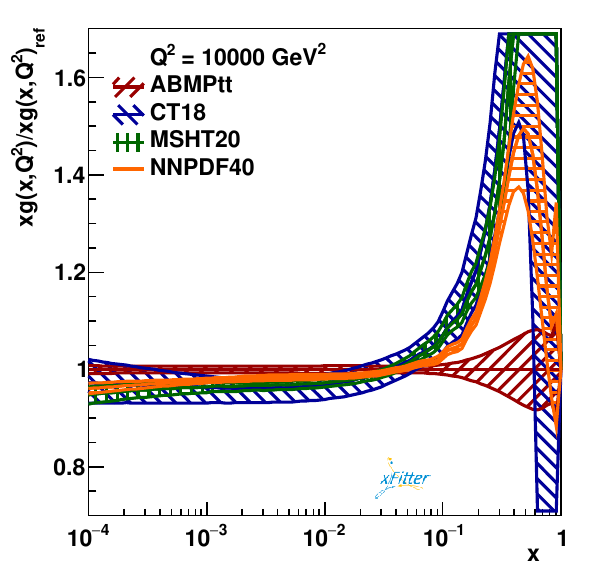}
       \caption{\label{fig:compagluon}
         Ratio of the gluon distributions in 
         recent global fits (CT18, MSHT20, NNPDF4.0)
         with respect to the  ABMPtt fit with $n_f=5$ at a scale Q=100 GeV. 
         See the text for more detail.}
\end{figure}

Finally, we provide a comparison of the ABMPtt PDFs to other PDF sets in Fig.~\ref{fig:compagluon}.
For il\-lu\-stra\-tive purposes, in Fig.~\ref{fig:compagluon} we compare our gluon distribution
at a scale $\mu = 100$~GeV to other modern gluon PDFs~\cite{Bailey:2020ooq, Hou:2019efy, Ablat:2023tiy,  NNPDF:2021njg}.
Fig.~\ref{fig:compagluon} shows that the gluon PDFs from CT18, MSHT20 and NNPDF4.0 are much larger for $x \gtrsim 5 \cdot 10^{-2}$, even by more then 50\% around $x \sim 0.5$,
while for $x \lesssim 10^{-2}$ all these gluon PDFs are smaller than ABMPtt by $\sim 5\%$. 
The uncertainty band of CT18~\footnote{The PDF uncertainties of the CT18 set, evaluated at 90\% confidence level, are rescaled to 68\% confidence level for consistency with other PDF sets.}
is mostly overlapping with the ABMPtt one in the range plotted. We verified that at large $x$ the central gluon distribution in the CT18 variants from Ref.~\cite{Ablat:2023tiy} are closer to our central gluon distribution than the original CT18 gluon~\footnote{We are grateful to Marco Guzzi and his collaborators for sharing with us the relevant PDF sets in
{\texttt{LHAPDF}} format.}.  The uncertainty on the gluon PDF is largest for CT18, followed by MSHT20, while those of ABMPtt and NNPDF4.0 are of similar size.
Differences in the gluon distribution by different PDF sets have been reported since early times, already for fits not yet including $t\bar{t} +X$ LHC data (see, e.g., the comparisons of $gg$ luminosities in Fig. 5 of the report~\cite{Alekhin:2011sk}). 
Differences qualitatively similar to those observed in Fig.~\ref{fig:compagluon} are illustrated, for example, in Fig.~22 of Ref.~\cite{Alekhin:2017kpj}, where older versions of these PDFs are shown, and their origins have been investigated and explained in previous studies
(see, e.g., Refs.~\cite{Accardi:2016ndt, Alekhin:2017kpj, Hou:2019efy, Bailey:2020ooq, Alekhin:2020edf, NNPDF:2021njg, Thorne:2014toa}).
We recall that the CT18 and NNPDF4.0 global fits
use particular input values 
for the on-shell top-quark mass $m_t^{\rm pole}$ and the strong coupling $\alpha_s(M_Z)$, which are not necessarily the same. Also, they are fixed before the fit, instead of being determined simultaneously with the PDFs. 
On the other hand, $m_t^{\rm pole}$ is treated as a nuisance parameter in the MSHT20 fit. These approaches do not capture all correlations between those parameters and the gluon PDF~\footnote{
  Besides providing their default PDF sets, 
the  CT18, MSHT20, and NNPDF4.0 collaborations perform fit variants, sometimes as dedicated studies. These variants either scan through fixed $\alpha_s(M_Z)$ values  or fit the optimal one, while allowing the gluon to vary each time
  (see e.g.  Ref.~\cite{Thorne:2014toa, Ball:2018iqk, Cridge:2021qfd}). 
}.
The differences in the gluon PDFs reported here (together with the extracted $\alpha_s(M_Z)$ value) have direct consequences in phenomenology, for example for the predicted cross section for Higgs boson production at the LHC (see e.g.~\cite{Accardi:2016ndt, Amoroso:2022eow}). 
More detailed interpretations and conclusions require coordination efforts among the different PDF collaborations, beyond the scope of this work.

\subsection{Top-quark mass}
\label{sec:mtop}

The simultaneous determination of the PDFs, $m_t(m_t)$ and $\alpha_s(M_Z)$ in the ABMPtt global fit leads to the following results. 
We obtain for the strong coupling in the $n_f=5$ flavor scheme at NNLO the  best-fit value 
\begin{eqnarray}
\label{eq:asMZ}
\alpha_s^{(n_f=5)}(M_Z)&=& 0.1150 \pm 0.0009 \, ,
\end{eqnarray}
which is very similar, both in terms of central value, and in terms of uncertainty, to the ABMP16 one $0.1147   \pm 0.0008 $. 
The latest PDG review~\cite{ParticleDataGroup:2024cfk} value is $\alpha_S(M_Z) = 0.1180 \pm 0.0009$, 
and we refer to the ABMP16 paper~\cite{Alekhin:2017kpj} for a thorough study of fit variants to explain the differences among those $\alpha_S(M_Z)$ determinations.

For the top-quark mass in the \msbar scheme~\cite{Langenfeld:2009wd} we obtain at NNLO the  best-fit value
\begin{eqnarray}
\label{eq:mt}
m_t(m_t) &=& 160.6 \pm 0.6~\GeV \, ,
\end{eqnarray}
which is slightly smaller, but also well compatible with the ABMP16 one $160.9 \pm 1.1$~GeV, and  with uncertainties reduced by a factor of approximately two. 
Eq.~(\ref{eq:mt}) corresponds to $m_t^{\rm pole} = 170.2 \pm 0.7$~GeV in the on-shell scheme, 
when performing a QCD conversion at three loops.
The PDG review~\cite{ParticleDataGroup:2024cfk} quotes from direct measurements a value of 
$m_t^{\mathrm{direct}} = 172.57 \pm 0.29\,\, \mbox{GeV}$, 
which depends on the definition of top-quark mass used in the Monte Carlo codes
adopted in the extraction. 
The precise relation of 
$m_t^{\mathrm{direct}}$ 
to $m_t(m_t)$ 
in Eq.~(\ref{eq:mt}) is currently an open problem, see e.g. Ref.~\cite{Hoang:2024zwl}.
The values of $m_t^{\rm{pole}} = 172.4 \pm 0.7\, \mbox{GeV}$ and $m_t(m_t) = 162.5^{+2.1}_{-1.5}\, \mbox{GeV}$ from sets of cross-section measurements quoted by the PDG align with our findings at a level of 1 to 2 $\sigma$.
However, these values are derived from a relatively small number of analyses/publications that do not cover the entire field and may not always reflect the most current data.
The charm- and bottom-quark masses $m_c(m_c)$ and $m_b(m_b)$ are almost identical to the ones obtained in the ABMP16 fit, since they are derived from the same input data, see also Appendix~\ref{appA:table}.

\begin{figure}[t!]
    \centerline{
      \includegraphics[width=0.6\textwidth]{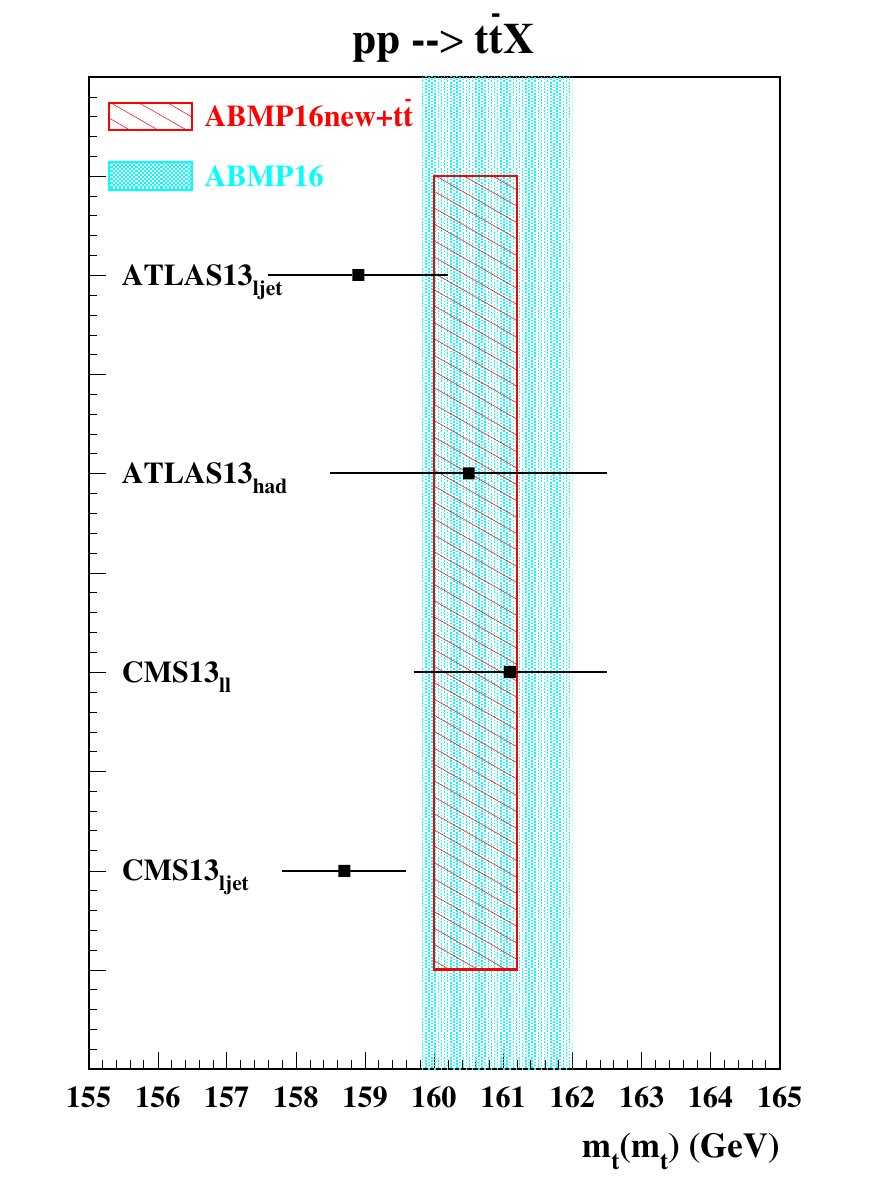}
}
\vspace*{-3mm}
\caption{
  \label{fig:mass}
The values of $m_t(m_t)$ obtained in the variants of present analysis including one single $t \bar{t}$ dataset (dots) compared to the band of present analysis (hatches) and the one of ABMP16 (shaded area).
}    
\end{figure}

In Fig.~\ref{fig:mass}, the two $m_t(m_t)$ values from ABMPtt and ABMP16 are depicted together with their uncertainty bands (the central values lie in the middle of the bands), and are further compared to the values of $m_t(m_t)$ from the variants of the ABMPtt analysis where only a single dataset with double-differential $t\bar{t}+X$ cross-section is considered at a time (see also Tab.~\ref{tab:tab3modified}).
The $m_t(m_t)$ values for all these analyses are compatible among each other. 
The analysis including only the CMS semileptonic dataset, that prefers a
slightly smaller $m_t(m_t)$ value  with respect to the ABMP16 and ABMPtt cases, is still compatible with the latter within 1.3 $\sigma$. 
\begin{figure}[h!]
  \includegraphics[width=0.7\textwidth]{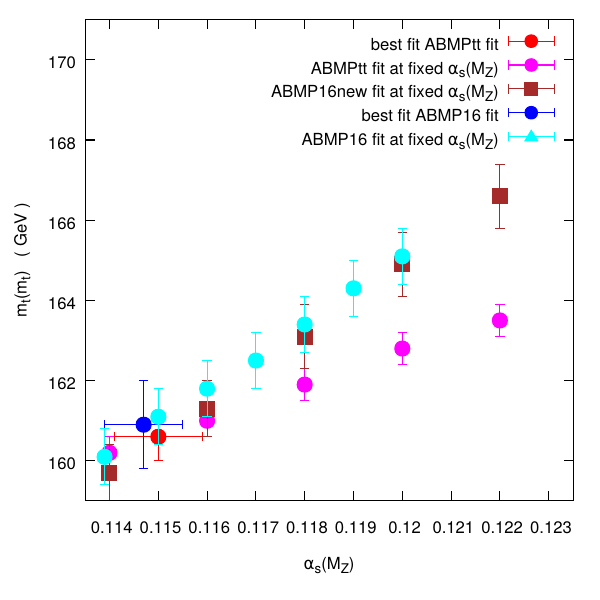}
\vspace*{-5mm}
  \caption{\label{fig:correl} Values of
    $m_t(m_t)$
    (extracted together with the PDFs)
    at fixed $\alpha_s(M_Z)$, for the ABMP16, the ABMP16new and the ABMPtt fits. In the case of ABMP16 and ABMPtt, the best fit values of $m_t(m_t)$ and $\alpha_s(M_Z)$  are also shown, considering the most general analyses, where also $\alpha_s(M_Z)$ is allowed to vary simultaneously with the PDFs and $m_t(m_t)$. 
  }
\end{figure}
As an illustration of the correlation between the $m_t(m_t)$ and $\alpha_s(M_Z)$ values, in Table~\ref{tab:mass} the results for the extraction of $m_t(m_t)$ using as input different fixed values of $\alpha_s(M_Z)$ are reported.
One can see an approximately 
linear increase of $m_t(m_t)$ with $\alpha_s(M_Z)$, with a correlation coefficient
decreased with respect to the ABMP16 and the ABMP16new fits, as shown in Fig.~\ref{fig:correl}. 
Altogether, our findings confirm that the double-differential
distributions considered in this analysis, differential
in $M({t\bar{t}})$ and in $y({t\bar{t}})$, are not par\-ti\-cu\-lar\-ly sensitive to $\alpha_s(M_Z)$.
To increase the sensitivity to the latter, one should use distributions differential
in the number of jets. For $t\bar{t} j$, $t\bar{t}jj$, $t\bar{t}jjj$, etc., however, theory predictions at NNLO accuracy are
not yet available. For the time being, the uncertainty on $\alpha_s(M_Z)$ is driven by correlations, considering that the scale evolution of quarks, gluons and the strong coupling $\alpha_s(\mu)$ are all coupled, and, as a consequence, there are a number of processes not involving top quarks, sensitive to $\alpha_s$ included in the fit. 
In particular, as even in the previous ABMP16 fit, DIS data have relevant impact on constraining $\alpha_s(M_Z)$ at low scales, whereas DY and other LHC data play a role at larger scale. 

\begin{table}[ht!]
\renewcommand{\arraystretch}{1.3}
\begin{center}                    
{\small                          
\begin{tabular}{|c|c|c|c|}   
\hline                                                    
\multicolumn{2}{|c|}{\multirow{2}{*}{$\alpha_s(M_Z,N_f=5)$}} & \multicolumn{2}{c|}{$m_t(m_t)$~(GeV)}  \\
\cline{3-4}                                                    
\multicolumn{2}{|c|}{} & ABMP16tt & ABMP16new
\\
\hline                                                    
Fitted & 0.1150(9) & 160.6(6) & $\,$ \\
\hline                                                    
\multirow{5}{*}{Fixed}& 0.114 & 160.2(4) & 159.7(7) \\
\cline{2-4}
   & 0.116 & 161.0(4) & 161.3(7) \\
\cline{2-4}
  & 0.118 & 161.9(4) & 163.1(8) \\
\cline{2-4}
  & 0.120 & 162.8(4) & 164.9(8) \\
\cline{2-4}
   & 0.122 & 163.5(4) & 166.6(8) \\
\hline
\end{tabular}
}
\caption{
\label{tab:mass}
\small 
The values of $m_t(m_t)$ obtained with different values of $\alpha_s(M_Z)$
for the ABMP16tt and ABMP16new fits.
}
\end{center}
\end{table}

\begin{figure}[h!]
    \centerline{
      \includegraphics[width=0.6\textwidth]{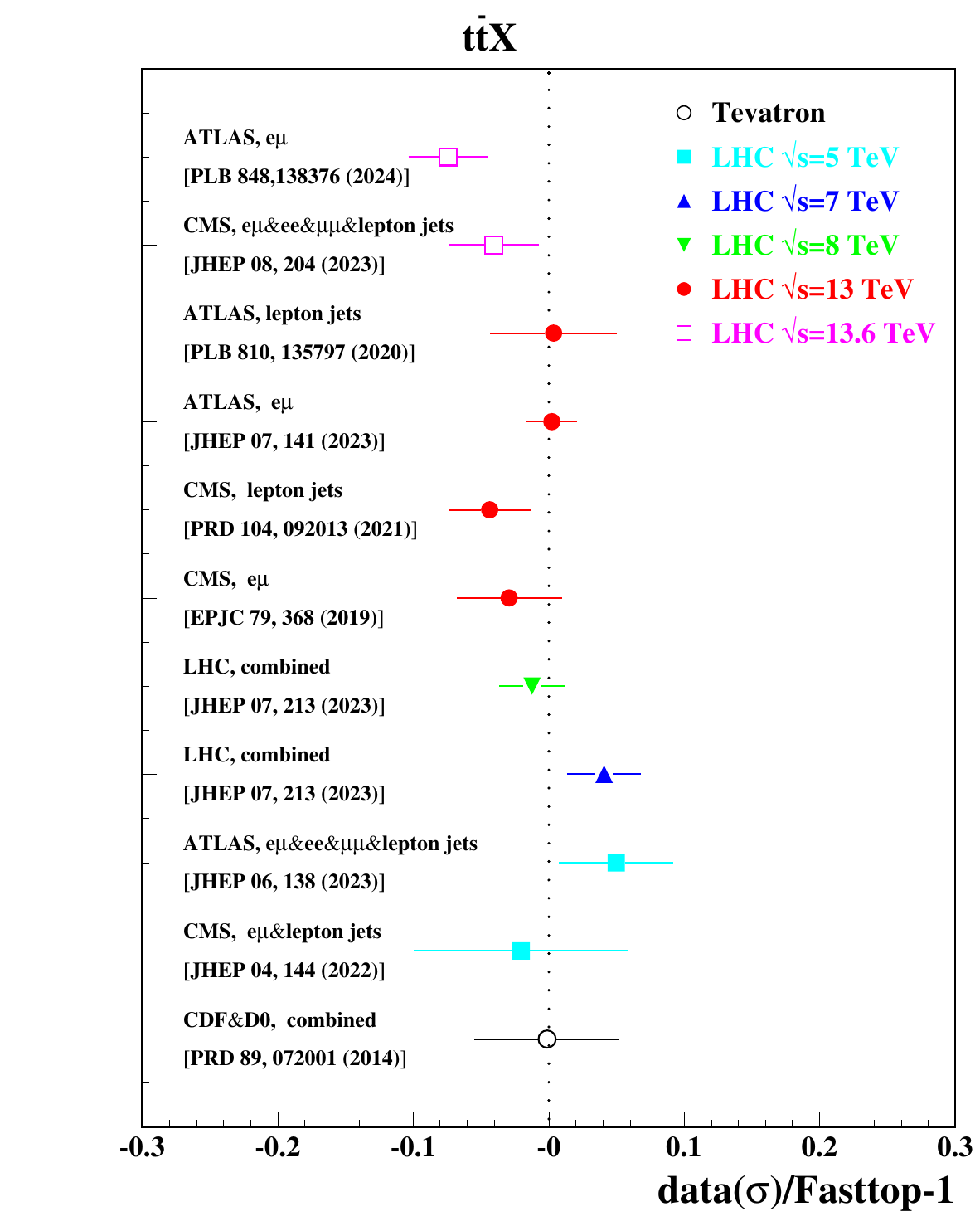}}
\vspace*{-3mm}
  \caption{\small
  \label{fig:totalcrosstt}
  Pulls obtained in present ABMPtt analysis for the inclusive total 
  cross sections 
  for $t\bar{t}+X$  production at Tevatron and various LHC energies, considering the
  selection of measurements  by the LHC Top Working Group~\cite{lhctopwg} (see text for more detail).
}  
\end{figure}
\begin{figure}[t!]
    \centerline{
      \includegraphics[width=0.6\textwidth]{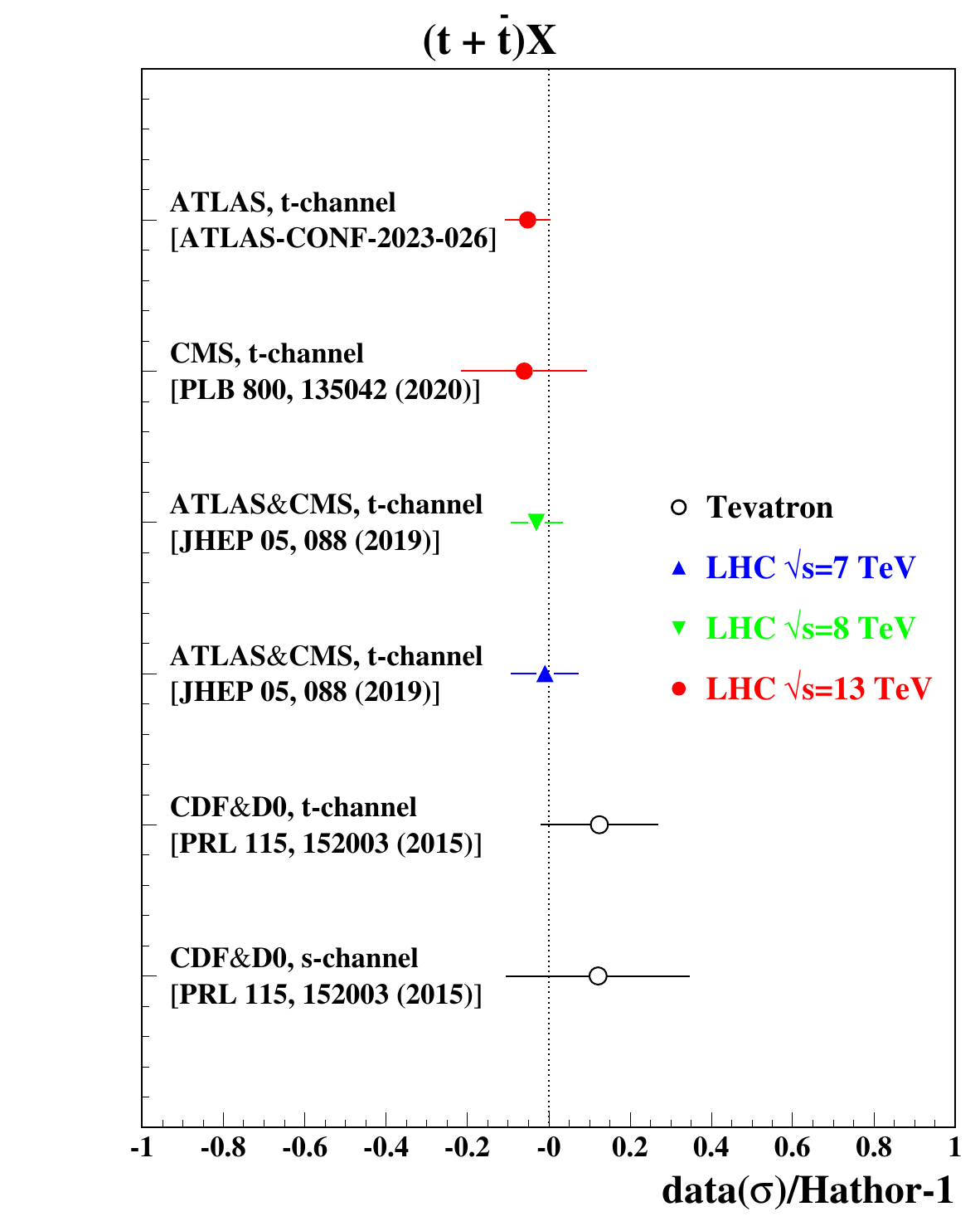}}
\vspace*{-3mm}
  \caption{\small
    \label{fig:totalcrosssingletop}
    Same as~Fig.~\ref{fig:totalcrosstt} 
    for the single-top inclusive cross sections at the TeVatron
    and the LHC, con\-si\-dering
    the selection of measurements by the {\it{LHCtopWG}}~\cite{lhctopwg} (see text for more detail).
  }  
\end{figure}

Finally, as a cross-check of the robustness of our results, 
the pulls of total cross sections for $t\bar{t}+X$ and ($t+X$) + ($\bar{t}+X$) production
at the Tevatron and the LHC, considering the most updated selection of measurements performed by the LHC Top Working Group, are shown in Figs.~\ref{fig:totalcrosstt} and~\ref{fig:totalcrosssingletop} (we recall that these data are also included in all 
the ABMPtt variants of the fits presented in Fig.~\ref{fig:gluon} 
and in Tables~\ref{tab:chi2modified} and~\ref{tab:mass}). 
In the case of $t\bar{t}+X$, agreement of theory predictions using as input the most general version of the
ABMPtt fit 
with the data is reached within $\sim$~2$\,\sigma$'s.
A simultaneous description of the combination of the experimental data from the ATLAS and CMS collaborations at $\sqrt{S}=$ 7 and 8 TeV is possible within 1.5 $\sigma$'s, 
pointing to some tension among the two.
Furthermore, it will be interesting to verify if future Run 3 data will confirm or not the results of the first Run 3 analyses considered in our fit.
On the other hand, in the case of single-top hadroproduction, an excellent agreement of our theory predictions with all considered datasets, including those with small uncertainty bands, is found, as evident from Fig.~\ref{fig:totalcrosssingletop}. 

\subsection{{\texttt{LHAPDF}} library}
\label{sec:lhapdf}
The ABMPtt PDFs are available as grids in the format of the {\texttt{LHAPDF}} library (version 6)~\cite{Buckley:2014ana} \footnote{
See \url{http://projects.hepforge.org/lhapdf}.}. 
For a fixed number of flavors, $n_f=3, 4$ and $5$, at NNLO, the grids are named
\begin{verbatim}
      ABMPtt_3_nnlo (0+29),
      ABMPtt_4_nnlo (0+29),
      ABMPtt_5_nnlo (0+29),
\end{verbatim}
and contain the central fit (set 0) together with additional 29 sets for the combined symmetric PDF uncertainties (including variations of the values of $\alpha_s$ and the heavy-quark masses). The latter correspond to $\pm 1\sigma$-variations.
The {\texttt{LHAPDF}} grids contain values of the heavy-quark masses $m_c(m_c)$, $m_b(m_b)$ and $m_t(m_t)$ 
in the \msbar scheme. 
They are correlated with the PDF parameters and they are different for each of the 29 PDF sets.
Appendix~\ref{appA:table} contains also values for the corresponding bottom- and the top-quark pole masses, 
$m_b^{\rm pole}$ and $m_t^{\rm pole}$.

The PDF sets for fixed $n_f=3, 4$ and $5$ flavors at NNLO use the corresponding strong couplings
$\alpha_s^{(n_f=3)}$, $\alpha_s^{(n_f=4)}$ and $\alpha_s^{(n_f=5)}$, which can be
related by the standard decoupling relations in QCD. 
The PDF set with three light-quark flavors {\tt ABMPtt\_3\_nnlo} is valid at all perturbative scales 
$\mu \gtrsim 1$~GeV, whereas the PDFs sets for $n_f=4$ or $5$, {\tt ABMPtt\_4\_nnlo} or {\tt ABMPtt\_5\_nnlo}, 
are only meaningful at scales which are much larger than the charm- or bottom-quark masses, where the decoupling is usually done. 
See also Ref.~\cite{Alekhin:2017kpj} for additional explanations.
  Additionally, we provide the $n_f = 5$ flavor NNLO PDF grid with the central value of $\alpha_S(M_Z)=0.118$ fixed, denoted as \texttt{ABMPttals118\_5\_nnlo} (0+29).

\subsection{{\texttt{PineAPPL}} interpolation tables}
\label{sec:grids}

Our \texttt{PineAPPL} interpolation tables are publicly available from the \texttt{Ploughshare} platform.\footnote{See \url{https://ploughshare.web.cern.ch/ploughshare/\#table1-w}.} Using these tables, one can compute theoretical predictions for $t\bar{t} + X$ production cross sections at hadron colliders in QCD at LO, NLO and NNLO (without K-factors), double-differential in $m(t\bar{t})$ and $y(t\bar{t})$, using as input whichever PDFs, $\alpha_S(M_Z)$ values, and renormalization and factorization scales multiple of $H_T /4$. They were generated
with the theoretical approach summarized in Section~\ref{sec:theo} 
 for Run 1 and Run 2 measurements by ATLAS and CMS, but can be potentially useful even for Run 3 and HL-LHC studies (and even at larger $\sqrt{S}$ energies), providing that the binning scheme and the kinematic range of the measurement is compatible. The tables are available for six different values of $m_t^{\rm{pole}}$ from $165$ to $177.5$~GeV with a step of $2.5$~GeV. They are named 
\begin{verbatim}
	xfitter-cms-ttbar-mt1650-arxiv-2108.02803,
	xfitter-cms-ttbar-mt1675-arxiv-2108.02803,
	xfitter-cms-ttbar-mt1700-arxiv-2108.02803,
	xfitter-cms-ttbar-mt1720-arxiv-2108.02803,
	xfitter-cms-ttbar-mt1750-arxiv-2108.02803,
	xfitter-cms-ttbar-mt1775-arxiv-2108.02803.
\end{verbatim}

In addition, we made public the tables with experimental $t\bar{t} + X$ data used in this and our previous work~\cite{Garzelli:2023rvx} as part of the \texttt{xFitter} project.\footnote{See \url{https://gitlab.cern.ch/fitters/xfitter}.} Together with ABMPtt PDFs in the format of the LHAPDF library, this allows the user to reproduce the description of the $t\bar{t} + X$ data from our fits.

\section{Conclusions and Outlook}
\label{sec:conclu}

We have shown that the use of state-of-the-art data on absolute 
total inclusive and normalized cross sections double differential in
$M({t\bar{t}})$ and $y({t\bar{t}})$ in a simultaneous NNLO fit of PDFs, $\alpha_s(M_Z)$ and $m_t(m_t)$
following the ABMP methodology, leads to a reduction of the uncertainties by a factor $\sim$~2
on both the gluon PDF and the $m_t(m_t)$ value, with respect to the previous ABMP16 fit. 
The latter was based on a much smaller set of data, using only $t\bar{t}+X$ and $(t+X)$~+~$(\bar{t}+X)$ total cross-sections data from Run 1 and Run 2 published up to 2016.
The value of $\alpha_s(M_Z)$ remains essentially unchanged, and its correlation with the top-quark mass is somewhat decreased. 
Improvements in the determination of $\alpha_s(M_Z)$ with LHC data would require a fit to data differential in the number of jets, 
considering the $\alpha_s$ dependence in the vertices for light-parton emission.
Thus, a simultaneous analysis of $t\bar{t}$ data and
jet data from hadroproduction and DIS, could be relevant in this respect and is within reach
although computationally extremely expensive and, therefore, beyond the scope of this paper, where we intentionally limited our focus to the impact of  top-quark related data on PDFs.
Moreover, NNLO theory predictions for $t\bar{t}$~+~one or more jets are not yet available.

While single-top total inclusive cross-section data from both the Tevatron and the LHC are all very well compatible with the present fit, the description of some of the $t\bar{t} + X$ datasets at the LHC looks more problematic. 
In particular the $l\!+\!j$ double-differential dataset by ATLAS turns out to be in slight tension with the all-hadronic one and with the dileptonic and semileptonic double-differential datasets by CMS. 
The combined ATLAS+CMS results for total inclusive $t\bar{t}$ cross sections at $\sqrt{S}=$~7 and 8~TeV point also to a slight tension among each other, while our theory description overestimates the new inclusive $t\bar{t}$ data at 13.6 TeV (Run 3). 
These tensions are in any case resolved within 2$\sigma$'s, and will benefit from the repetition of the Run~2 and Run~3 analyses with an increased integrated luminosity. 
On the other hand, the $l\!+\!j$ double-differential data by CMS, corresponding to a much  larger integrated luminosity, are well described by the theory and, having small uncertainties,  drive   our fit.
We emphasize that efforts by the ATLAS and CMS collaborations to combine their differential measurements in different channels would indeed be very useful and help in better understanding systematics and correlations, as well as with discovering  possible issues that  might affect some of the datasets.      

We emphasize that our NNLO computations, using the \texttt{MATRIX+PineAPPL} framework, are exact, not involving any approximation or any $K$-factor, often used in NNLO fits  notwithstanding the risk of distortion of predictions for differential distributions.
However, considering that the $q_T$-subtraction method,     used in \texttt{MATRIX}
for canceling infrared divergences,                 is   non-local,       we are         neglecting the possible effects of consequent power corrections on our predictions, that we have estimated of the order of $\lesssim 1\%$ on the basis of comparisons with predictions using the STRIPPER local subtraction method. 

This work made use of top-quark data collected by ATLAS and CMS in a limited rapidity range, up to $|y({t\bar{t}})| < 2.5 $, which are extrapolated to the full
phase space in order to determine total inclusive cross sections.
The uncertainties related to this extrapolation are included in the errorbars of the experimental datapoints.
As a future study, it will be interesting to include in our analysis even LHCb data, which probe the region $2 < y({t\bar{t}}) < 4.5$, that for the time being are however only available with fiducial cuts. This can, potentially, both verify the correctness of these extrapolations, and constrain the gluon distribution at even larger $x$ values than currently accessed.
So far, the LHCb collaboration has provided analyses of $t\bar{t} + X$ events in Refs.~\cite{LHCb:2015nta, LHCb:2018usb}, however, without performing an extrapolation to the full phase-space. 
 
We have assumed that the unfolding procedure to reconstruct top quarks implemented in the experimental analyses used as a basis of this work is independent from the definition of top-quark mass in the Monte Carlo codes, $m_t^{\mathrm{MC}}$. 
This requires that the experimental collaborations provide robust estimates for all possible uncertainties affecting the unfolding procedure, which are included in the quoted uncertainty for the top-quark mass.
On the other hand, the relation between $m_t^{\mathrm{MC}}$ and the top-quark mass in well defined renormalization schemes, such as the on-shell or the \msbar scheme, is still topic  of study (see e.g.~\cite{Hoang:2024zwl} and references therein). 
Thus, the results are subject to additional uncertainties related to the difference between $m_t^{\mathrm{MC}}$ and $m_t(m_t)$,
which cannot be fully quantified at present. 
It is of paramount importance to address this topic with further improved theory predictions,
in order to scrutinize the pressing issue of stability of the electroweak vacuum. 
The extracted value of $m_t^{\rm pole}$ points towards vacuum stability even at very high scales, see e.g. Refs.~\cite{Alekhin:2012py,Buttazzo:2013uya,Bednyakov:2015sca}.

Follow-ups of this work may involve the use of improved theory predictions, including corrections due to the resummation of logarithms associated to soft-gluon production close to threshold and below it (Coulomb logarithms) 
(see, e.g., Ref.~\cite{Kiyo:2008bv} for the evaluation of next-to-leading logarithmic contributions to the
$M({t\bar{t}})$ 
  distribution  arising from $t\bar{t}$ color-singlet and color-octet
  confi\-gu\-rations in non-relativistic QCD). 
  These corrections, mainly affecting the top-quark mass value, are not particularly important for our present fit, since the  $M({t\bar{t}})$ bins close to threshold are large enough. 
On the other hand, they might become particularly relevant when the size of the $M({t\bar{t}})$ bins considered in the experimental analyses will become smaller, as plausible  in the case of future high-luminosity data. 
In that case, it will be preferrable to fit some short distance mass or $m_t^{\rm pole}$ instead of $m_t(m_t)$, considering the behavior of the latter very close to threshold (see e.g.~\cite{Dowling:2013baa,Garzelli:2020fmd}).

\subsection*{Acknowledgements}
We are grateful to Javier Mazzitelli for continuous assistance with the \texttt{MATRIX} code and for having provided us a special version of it, tailored to our needs. 
The work of S.A. and S.-O.M. has been supported in part by the Deutsche Forschungsgemeinschaft through the Research Unit FOR 2926, {\it Next  Generation pQCD for Hadron Structure: Preparing for the EIC}, project number 40824754.
The work of M.V.G. and S.-O.M. has been supported in part by the Bundesministerium f\"ur Bildung und Forschung under contract 05H21GUCCA.
The work of O.~Z. has been supported by the {\it Philipp Schwartz Initiative} of the Alexander von Humboldt foundation. 

\appendix
\section{Parameters of \textrm{ABMPtt} PDFs}
\label{appA:table}

In Table~\ref{tab:parameters} we list the values of $\alpha_s(M_Z)$ and the heavy-quark masses emerging from their simultaneous fit with the ABMPtt PDFs for each of the eigenvectors. 
The values of the bottom- and top-quark masses in the on-shell scheme from the conversion of the masses in the \msbar scheme at three-loops, $m_b(m_b)$ and $m_t(m_t)$, using {\texttt{RunDec}}~\cite{Chetyrkin:2000yt}, are also provided.
Electroweak effects, known up to two-loops and recently considered in Ref.~\cite{Kataev:2022dua}, are neglected in this conversion.

\begin{table}[ht!]
\begin{center}
\begin{tabular}{|l|c|c|c|c|c|c|c|}
\hline
PDF set
  & $\alpha_s^{(n_f=3)}(M_Z)$
  & $\alpha_s^{(n_f=5)}(M_Z)$
  & $m_c(m_c)$~[GeV]
  & $m_b(m_b)$~[GeV]
  & $m_b^{\rm pole}$~[GeV]
  & $m_t(m_t)$~[GeV]
  & $m_t^{\rm pole}$~[GeV]
\\[0.5ex]
\hline
    {\bf \phantom{0}0} &  {\bf 0.10309}   &   {\bf 0.11498} &   {\bf 1.258} &   {\bf 3.790} &   {\bf 4.485} &   {\bf 160.63} &   {\bf 170.15} 
\\[0.5ex]
\hline
\phantom{0}1  &   0.10309&   0.11498&     1.258&     3.790&     4.485&    160.63&    170.15\\
\phantom{0}2  &   0.10310&   0.11499&     1.258&     3.790&     4.485&    160.63&    170.16\\
\phantom{0}3  &   0.10309&   0.11498&     1.258&     3.790&     4.485&    160.63&    170.15\\
\phantom{0}4  &   0.10312&   0.11502&     1.258&     3.790&     4.485&    160.63&    170.16\\
\phantom{0}5  &   0.10316&   0.11507&     1.258&     3.790&     4.486&    160.63&    170.16\\
\phantom{0}6  &   0.10309&   0.11499&     1.258&     3.790&     4.485&    160.63&    170.15\\
\phantom{0}7  &   0.10306&   0.11494&     1.258&     3.790&     4.484&    160.63&    170.15\\
\phantom{0}8  &   0.10294&   0.11478&     1.258&     3.790&     4.481&    160.63&    170.13\\
\phantom{0}9  &   0.10332&   0.11529&     1.258&     3.790&     4.491&    160.63&    170.18\\
10            &   0.10289&   0.11473&     1.258&     3.790&     4.479&    160.63&    170.13\\
11            &   0.10314&   0.11505&     1.256&     3.790&     4.486&    160.63&    170.16\\
12            &   0.10313&   0.11504&     1.257&     3.790&     4.486&    160.63&    170.16\\
13            &   0.10316&   0.11504&     1.274&     3.790&     4.486&    160.63&    170.16\\
14            &   0.10315&   0.11506&     1.255&     3.790&     4.486&    160.63&    170.16\\
15            &   0.10327&   0.11522&     1.257&     3.789&     4.489&    160.62&    170.17\\
16            &   0.10307&   0.11496&     1.259&     3.789&     4.483&    160.63&    170.15\\
17            &   0.10315&   0.11500&     1.259&     3.888&     4.592&    160.62&    170.15\\
18            &   0.10313&   0.11499&     1.258&     3.861&     4.562&    160.65&    170.18\\
19            &   0.10310&   0.11500&     1.259&     3.789&     4.484&    160.61&    170.13\\
20            &   0.10313&   0.11503&     1.257&     3.800&     4.497&    160.67&    170.20\\
21            &   0.10313&   0.11503&     1.257&     3.794&     4.490&    160.59&    170.12\\
22            &   0.10332&   0.11528&     1.256&     3.791&     4.492&    160.84&    170.40\\
23            &   0.10323&   0.11515&     1.261&     3.791&     4.489&    160.66&    170.20\\
24            &   0.10325&   0.11519&     1.258&     3.798&     4.498&    160.71&    170.26\\
25            &   0.10303&   0.11490&     1.255&     3.799&     4.493&    160.64&    170.16\\
26            &   0.10309&   0.11499&     1.260&     3.788&     4.483&    160.55&    170.07\\
27            &   0.10321&   0.11513&     1.257&     3.792&     4.490&    160.66&    170.20\\
28            &   0.10343&   0.11543&     1.256&     3.796&     4.500&    161.13&    170.72\\
29            &   0.10321&   0.11514&     1.258&     3.787&     4.485&    160.79&    170.34\\[0.5ex]
\hline
\end{tabular}
\caption{\small 
  \label{tab:parameters}
  Values of $\alpha_s^{(n_f=3)}(M_Z)$,  $\alpha_s^{(n_f=5)}(M_Z)$ and heavy-quark masses 
  $m_c(m_c)$, $m_b(m_b)$ and $m_t(m_t)$ 
  in the \msbar scheme obtained for the individual eigenvectors of the ABMP16tt fit. 
  For bottom and top, also the values for pole masses $m_b^{\rm pole}$ and
  $m_t^{\rm pole}$ in the on-shell scheme are given.
}
\end{center}
\end{table}

\section{ABMPtt vs. ABMP16new vs. ABMP16 PDFs}
\label{appB:allpdfs}
\begin{figure}[h!]
  \begin{center}
  \includegraphics[width=0.48\textwidth]{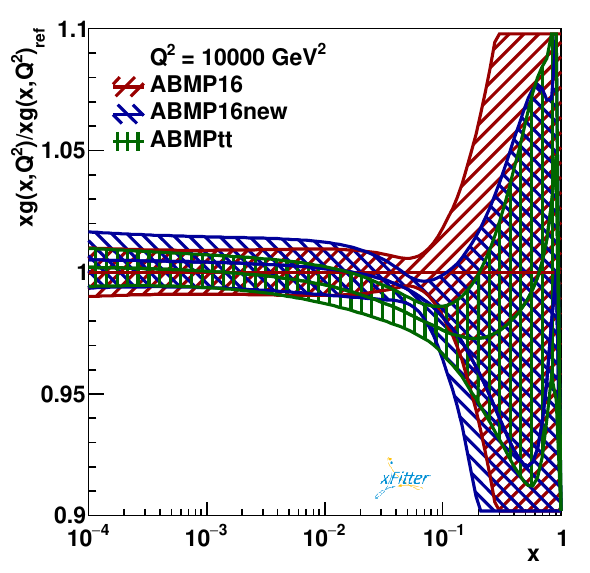}
  \includegraphics[width=0.48\textwidth]{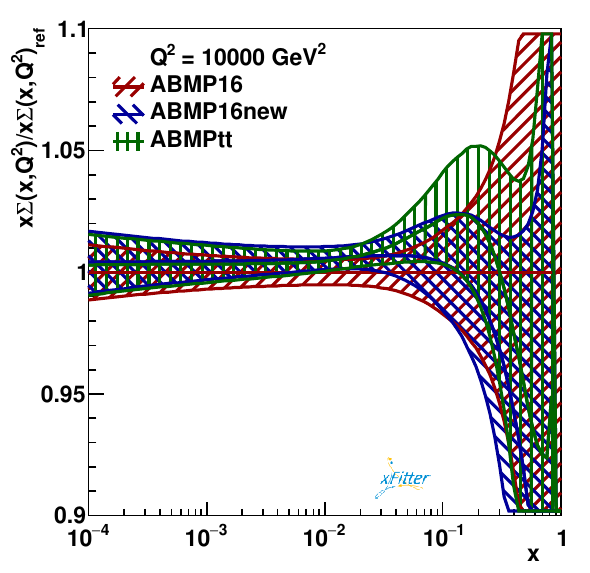}
  \includegraphics[width=0.48\textwidth]{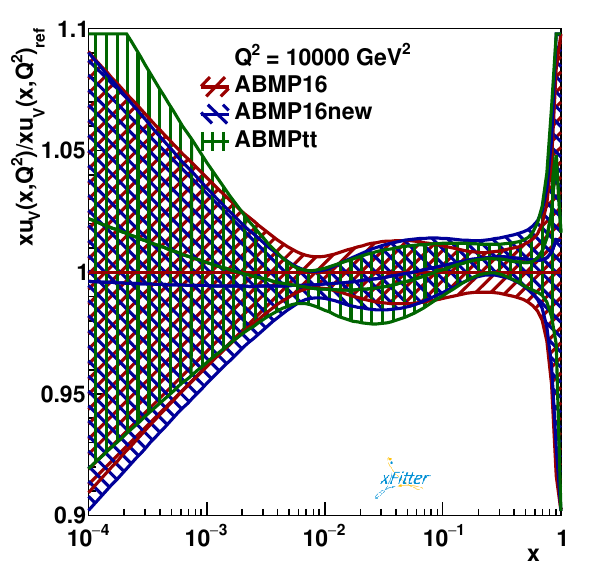}
  \includegraphics[width=0.48\textwidth]{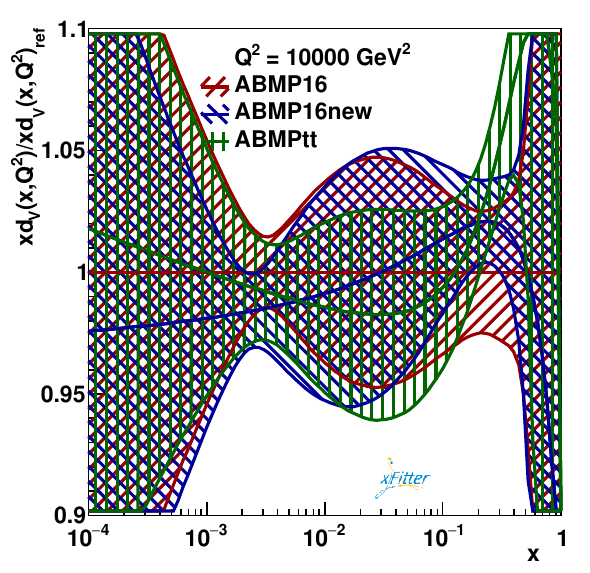}
       \caption{\label{fig:compa16vs16newvstt_nf=5}
         Ratio of the gluon, sea-quark, valence up- and down-quark distributions in the  ABMPtt and ABMP16new fits with respect to the published ABMP16 PDFs at a scale $Q=100$~GeV
         for $n_f=5$.}
       \end{center}
       \end{figure}

A comparison of the PDFs for gluon, sea and valence quarks in the ABMPtt, ABMP16new and ABMP16 fit for $n_f=5$ and $Q=100$~GeV is shown in Fig.~\ref{fig:compa16vs16newvstt_nf=5}.

\bibliographystyle{JHEP}
\bibliography{ttab2} 
\end{document}